\documentclass{aa}  

\usepackage{graphicx}
\usepackage{natbib}
\defcitealias{2020A&A...641A..83K}{Paper~I}
\usepackage{txfonts}
\usepackage{amssymb}
\usepackage{soul}
\usepackage{cancel}
\usepackage{longtable}
\usepackage{ltablex}
\usepackage{multicol}
\usepackage{supertabular}
\usepackage{xcolor}

\usepackage{hyperref}
\hypersetup{
    colorlinks=true,
    citecolor = blue,
    linkcolor=blue
            }
\begin{document}

   \title{What powers the most energetic flares in the \emph{Kepler} field?}

   \subtitle{Revisiting the superflaring star KIC\,2852961}

\titlerunning{}

\author{Zs.~K\H{o}v\'ari\inst{1,2}
        \and
          B.~Seli\inst{1,2}
        \and
          L.~Kriskovics\inst{1,2}
        \and
          K.~Ol\'ah\inst{1,2}
        \and
          P.~Odert\inst{3}
        \and
          M.~Leitzinger\inst{3}
        \and
          B.~Schmercz\inst{1,4}
        \and
          K.~Vida\inst{1,2}
        \and
          R.~Greimel\inst{5}
        \and
          B.~Cseh\inst{1,2}
        \and
          A.~G\"orgei\inst{1,2}
        \and
          T.~Granzer\inst{6}
        \and
          P.~S\'agi\inst{1,2}
        \and
          M.~Weber\inst{6}
  }

   \institute{Konkoly Observatory, HUN-REN Research Centre for Astronomy and Earth Sciences, Konkoly Thege \'ut 15-17., H-1121 Budapest, Hungary\\
        \email{kovari@konkoly.hu}
        \and
            HUN-REN CSFK, MTA Centre of Excellence,  Konkoly Thege \'ut 15-17., H-1121 Budapest, Hungary
        \and
           Institute of Physics/IGAM, University of Graz, Universit\"atsplatz 5, A-8010 Graz, Austria
        \and
           E\"otv\"os Lor\'and University, P\'azm\'any P\'eter s\'et\'any 1/A, H-1117 Budapest, Hungary 
        \and
           RG Science, Graz, Austria
        \and
            Leibniz-Institute for Astrophysics Potsdam (AIP), An der Sternwarte 16, D-14482 Potsdam, Germany
            }

   \date{Received ...; accepted ...}       


 
  \abstract
   {Flare activity in red giants, once thought to be extremely rare, is actually quite common when continuously monitored with high-precision space photometry. In fact, the most powerful flares, also known as "superflares", with bolometric energies exceeding 10$^{38}$\,erg have been observed on red giant stars.}
   {KIC\,2852961 is a red giant star producing the highest-energy flares in the \emph{Kepler} field, and is also known for its large-amplitude starspot-induced light variations. Our goal is to find out what causes the extreme flare activity and flare energies in this star.}
   {For our investigations, we use ground-based spectroscopic measurements and space photometric data, which we process using standard tools. We also collect relevant data from databases of large surveys such as Gaia.}
   {Our new radial velocity data confirm that the red giant star is part of a close binary system, in which tidal forces play a crucial role in controlling the evolution of the binary system. Our results show that tidal evolution has driven the system close to orbital circularization and rotational synchronization, but it did not completely constrain the differential rotation of the red giant, in which the magnetic dynamo can operate extremely efficiently. We find that high-energy (super)flares are primarily controlled by the internal magnetic activity of the active giant component, rather than phase-dependent interactions with the secondary star.}
   {The large spot-induced rotational light modulations indicate extensive starspots that can store enormous amounts of magnetic energy. This provides the conditions for the formation of extremely large flares, confirming the close relationship between large spot coverage and high flare energies and/or frequency. Behind all this, differential rotation may play a dual role in the flare activity of KIC\,2852961: it contributes both to the formation of large starspots and to the shearing and twisting of coronal magnetic fields that can power energetic superflares.}


   \maketitle
%

\section{Introduction}

The most powerful flares (also known as "superflares") with bolometric energies reaching $10^{38}$-$10^{39}$\,erg have been observed on red giant stars \citep[e.g.][]{2021A&A...647A..62O,2025ApJ...980..268M}. However, flares on giants are inherently more difficult to detect with ground-based telescopes due to the high background brightness of the stellar surface. Moreover, compared to observations from space, ground-based observations are less likely to detect flares due to lower accuracy and atmospheric noise, as well as a lower cadence. The study of the flare activity of red giants has been given a significant boost by recent and current space missions such as CoRoT \citep[e.g.][]{2022AJ....164..223R}, \emph{Kepler} and K2 \citep[e.g.][]{2016ApJ...829...23D,2017ApJS..232...26V,2021A&A...645A..42I}, or the still ongoing TESS \citep[e.g.][etc.]{2020AJ....159...60G,2023A&A...669A..15Y,2025A&A...694A.161S}. 
These surveys have revealed that flare activity in red giants, once thought to be extremely rare, is actually detectable and quite common when continuously monitored with high precision. The uninterrupted coverage and sub-mmag precision enable not only the detection of rare, short-lived flares, but also the investigation of magnetic variability, rotational modulation and activity cycles, ultimately the quantitative characterization of stellar magnetism and evolution.

The operation of the magnetic dynamo accumulates magnetic energy in the stellar atmosphere through the interaction between turbulent plasma motions and plasma shear, which is released during large flares, thereby, as a negative feedback, confining the large scale magnetic field strength.
However, red giants present a markedly different stellar environment compared to their main-sequence counterparts. Characterized by deep convective envelopes, low surface gravities, and relatively slow rotation rates, red giants host magnetic fields that are more spatially diffuse and therefore weaker \citep{2015A&A...574A..90A} compared to rapidly rotating active dwarfs \citep{2024ApJ...971..112W}. These factors significantly influence the structure of the outer atmosphere, particularly the corona where magnetic reconnections (i.e. flares) occur, and ultimately, the efficiency of the stellar dynamo. Although the large stellar radii of the red giants imply the potential for expansive magnetic loops and large-scale magnetic structures, the total magnetic energy available for flare production is constrained by both the strength and topology of the magnetic field and the physical conditions of the coronal plasma. 
It is an interesting question whether there is, and if so, what is, the maximum flare energy that can be released at a time on a given star \citep[see the recent study on the most energetic stellar flare so far by][]{2024ApJ...977....6G}. Either way, it is likely that there may be a kind of scaling effect behind the generation of flares in the sense that the greater the magnetic activity, the greater the total magnetic energy emitted by the flares \citep{2020A&A...641A..83K}.
In conclusion, it is certain that further study of the energetic characteristics of flare events, including superflares in evolved stars, and how they relate to other activity phenomena, may lead us closer to understanding the nature of superflares observed either on red giants or on solar-type stars, which becomes particularly relevant when considering the star closest to us.

In our previous study \citep[][Paper~I from now on]{2020A&A...641A..83K}, we analyzed the flaring activity of a specific target from the \emph{Kepler} Input Catalog (KIC) under the entry-name KIC\,2852961 (see also as 2MASS~J19261136+3803107, TIC\,137220334) using space telescope data from \emph{Kepler} and the then-available TESS Sector 14. The star caught our attention during a systematic search for flaring \emph{Kepler} giants \citep{2021A&A...647A..62O} by producing exceptionally high flare energies. In the long-cadence \emph{Kepler} light curve, we observed rotational variability, which changes from one rotation to the next, typical of stars with constantly renewing spotted surfaces. In addition, 59 flare events were present, including very large ones. Careful examination of the light curve revealed that the integrated total flare energy for a given period correlates with the average amplitude of the rotational modulation caused by starspots---a very significant finding, supporting the assumption that (at least in first approximation) differences in flare energies of different luminosity class targets are very likely due to size effect.

In recent years, a significant amount of new TESS observations of the star have become available, which we have supplemented with radial velocity measurements, which provide a sufficient basis for an orbital solution. This work includes the processing of ground-based and space-based observations collected since the publication of \citetalias{2020A&A...641A..83K}, as follows. In Sect.\,\ref{Sect_obs} we present the data to be used. In Sect.\,\ref{Sect_astropars}, we summarize what we know about the astrophysical characteristics of KIC\,2852961. The methods used and the results derived from them are presented in Sect.\,\ref{Sect_metandres}.
Discussions of the findings are provided in Sect.\,\ref{Sect_disc}, and a summary with our conclusions can be found in Sect.\,\ref{Sect_sum}.

\section{Observations}\label{Sect_obs}

\subsection{Spectroscopic observations from Konkoly Observatory}

Most of our spectroscopic data were collected at Piszkéstető Mountain Station of Konkoly Observatory, Hungary during April-May 2020 and April 2025-Aug 2026 with the $R=21000$ fiber-fed échelle spectrograph mounted on the 1-m Ritchey-Chrétien-Coudé (RCC) telescope. The practical wavelength range of the spectrograph is 4200-7900\,\AA. Exposure time was usually one hour with 15 exceptions, when only half of this was possible due to bad weather. The average signal-to-noise (S/N) ratio was 48, while the minimum was 10 and the maximum was 114. Although we have already used the 2020 data averaged per night for radial velocity (hereafter RV) measurements in \citetalias{2020A&A...641A..83K}, we will now use the measurements again, re-reducing them one by one, to obtain a complete RV curve with the new data. Ultimately, we were able to use a total of 101 spectra from the observations.

Data reduction for the Piszkéstető data was carried out with \texttt{piszkespipe}\footnote{\url{https://pypi.org/project/piszkespipe}}, a heavily modified version of CERES \citep{piszkespipe} optimized for the Piszkéstető spectrograph. The standard reduction steps include zero-level and dark current removal, flat fielding,  removal of the Blaze-curvature, dispersion calibration using a Th-Ar calibration lamp and continuum normalization.

Radial velocities were determined by computing the cross-correlation function (CCF) in respect to synthetic template spectra. As a preliminary step, a CCF mask was applied, which assigns a weight of 1.0 to narrow spectral lines and zero weight to broader features. Uncertainties are estimated from the width of the CCF peak and the signal-to-noise ratio close to the Mg\,{\sc i} triplet ($\lambda\approx5130$\,\AA). For more details, see \cite{piszkespipe}. The average radial velocity offset of HD\,159222, measured as an RV standard \citep[$-$51.603\,km\,s$^{-1}$,][]{2018A&A...616A...7S}, was +0.744\,km\,s$^{-1}$ for the 2020 data, which changed to +0.520\,km\,s$^{-1}$ for the subsequent measurements (due to a CCD replacement). All of these corrections have already been taken into account in the final RV values from Piszkéstető, which are given in Table\,\ref{Table:piszkes_RV}.

\subsection{STELLA spectroscopy}

Twenty-five more spectroscopic observations were made with the fully robotic 1.2-m telescope STELLA at Iza\~na Observatory, Tenerife, Spain \citep{2010AdAst2010E..19S} between 11 November, 2024 and 12 Aug 2026. The telescope is equipped with the fixed-format $R=55000$ STELLA échelle spectrograph (SES) with an e2v 4k$\times$4k CCD featuring 15\,$\mu$m pixels and a fiber-fed structure. The wavelength coverage ranges from 3900 to 8700\AA. Typical observations used an integration time of 1800\,s, yielding an average S/N ratio of about 29. The SES data reduction pipeline is based on IRAF and is described in detail in \citet{2012SPIE.8451E..0KW}. For further details on the reduction process, we refer to \citet{2016SPIE.9910E..0NW}.

From the STELLA-SES spectra relative radial velocities are determined from cross correlation with synthetic spectra calculated using the Turbospectrum code \citep{turbo}, based on MARCS model atmospheres \citep{2008A&A...486..951G} and the Vienna Atomic Line Database \citep[VALD-3,][]{vald3} line list.
Absolute velocities are obtained from relative velocities by eliminating the instrumental offset with a zero-point shift of +0.503\,km\,s$^{-1}$ \citep[see][]{2012AN....333..663S}. The RV observations from STELLA are listed in Table\,\ref{Table:STELLA_RV}.

\subsection{Space photometry from \emph{Kepler} and TESS}

A total of 19 \emph{Kepler} light curves are available from our target, of which 18 are long-cadence time series from quarters Q0-Q17 with a 30-minute time resolution (effective integration time), and one short-cadence light curve from Q4 with a one-minute sampling. The full \emph{Kepler} data of KIC\,2852961 are presented in more detail in \citetalias{2020A&A...641A..83K}. As a further step forward, in Sect.\,\ref{Sect_SML}, we attempt to fit a time-series spot model to the full \emph{Kepler} data set.

So far, TESS observed KIC\,2852961 in the following 27-day long sectors: 14, 40, 41, 54, 55 and 74. The observing cadence was 1800\,s in Sector 14, 600\,s in Sectors 40-41-54-55, and 200\,s in Sector 74. For the TESS photometric reduction the Quick Look Pipeline \citep{2020RNAAS...4..204H} was used. The final form of the TESS light curves were obtained by stitching together half-sector long segments using constant shifts, to ensure smoothness between the segments. Unfortunately, the 27-day length of the sectors does not allow for the coverage of full rotation periods ($\sim$35\,d) of our target, therefore, unlike the \emph{Kepler} data, the use of TESS light curves is limited to the detection of flares and the estimation of the spot modulation amplitudes.

\section{Astrophysical data of KIC\,2852961}\label{Sect_astropars}

\subsection{Astrophysical parameters from spectral synthesis}
We carried out a detailed spectroscopic analysis based on spectral synthesis using the code SME \citep{piskunov_sme} and MARCS atmospheric models \citep{gustafsson_marcs} on both the Piszkéstető dataset using the ten best quality spectra with S/N>60. STELLA spectra were not used in the spectral synthesis due to their relatively low S/N values. Atomic line parameters were taken from the VALD database \citep{kupka_vald}. Macroturbulence was estimated to be $v_{\mathrm{mac}}=5.2\,\rm{km\,s^{-1}}$ using the equation from \citet[][see their Eq.\,4]{valenti_macro}. We note that the method described in \cite{valenti_macro} was calibrated for dwarfs and subgiants, but KIC\,2852961 is a K giant and its line broadening is rotationally dominated, hence,  the precision of macroturbulence is not expected to have a strong effect on the derived astrophysical parameters. Nevertheless, we ran the same synthesis loop with $v_{\rm mac}$=3.2\,km\,s$^{-1}$ as well, and the difference was less than 1\% for every astrophysical parameters.  
The values of $T_{\rm eff}$, $\log g$, [Fe/H], $v\sin i$ and $v_{\rm mic}$ obtained with our iterative method \citep{kriskovics2019} are listed in Table\,\ref{tab1}.

\subsection{Distance, effective temperature, radius}\label{sect:stellarparams}
The improved Gaia DR-3 parallax of 1.2887±0.0106\,mas corresponds to a distance of $776.0^{+6.4}_{-6.3}$\,pc, which is a $\sim$5\% decrease compared to the distance given in \citetalias{2020A&A...641A..83K} (based on Gaia DR-2). 
Interstellar extinction $A_V$=$0\fm233$ is derived from the Bayestar 3D interstellar dustmaps\footnote{\url{https://github.com/gregreen/dustmaps}}  \citep{2018MNRAS.478..651G}.
Adopting $V_{\rm max}$=$10\fm30$ from \citetalias{2020A&A...641A..83K} and substituting the above values into the distance modulus formula, we get $M_V$=$0\fm618\pm0\fm018$ for the absolute visual brightness. Taking $BC_V$=$-0\fm454\pm0\fm009$ from \citet[][see their Eq.18]{1999A&AS..140..261A} yields a bolometric magnitude of $M_{\rm bol}$=$0\fm164\pm0\fm020$, convertible to $L=67.7^{+1.3}_{-1.2}\,L_{\odot}$. We note that this improved luminosity value, resulting from the reduction in distance value, is a $\sim$12\%  decrease in luminosity compared to the value derived from Gaia DR-2 in \citetalias{2020A&A...641A..83K}. The Stefan-Boltzmann law, assuming $T_{\rm eff}$=4760$\pm$20\,K, gives a radius of $R=12.10\pm0.22R_{\odot}$.

 \begin{center}
 \begin{table}[thb]
\caption{Astrophysical parameters of KIC\,2852961} \label{tab1}
 \begin{tabular}{lcc}
 \hline\hline\noalign{\smallskip}
  Parameter               &  Value & Source\\
  \hline\noalign{\smallskip}
  Primary spectral type            & K0III  &  this paper \\
  Distance [pc] &  $776^{+6.4}_{-6.3}$ & Gaia DR-3\\
  $V_{\rm max}$ [mag] & $\approx$$10.30$ & {\citetalias{2020A&A...641A..83K}} \\
  $M_{\rm bol}$     [mag]    & $0.164\pm0.020$ & this paper\\
  Luminosity [${L_{\odot}}$]     & $67.7^{+1.3}_{-1.2}$ & this paper  \\
  $T_{\rm eff}$ [K] &           $4760\pm20$ & this paper   \\
  Radius [$R_{\odot}$] &      $12.10\pm0.22$  & this paper  \\
  Mass [$M_{\odot}$]   & $1.54\pm0.20$ & this paper  \\
  $\log g$ [cgs]  &   $ 2.52\pm0.09$ & this paper \\
  Metallicity [Fe/H] &  $-0.28\pm0.08$ & this paper \\
  $v\sin{i}$ [kms$^{-1}$]  &   $17.5\pm0.7$ & this paper \\
  $v_{\rm mic}$ [km\,s$^{-1}$] & $1.34\pm0.1$ & this paper \\
  Inclination  & $69^{\circ}\lesssim i \lesssim 77^{\circ}$ & this paper \\
  $P_{\rm rot}$ [d]  &  $35.515^{+0.454}_{-0.226}$ & this paper \\
  Binary separation $a$ [AU] & $\approx 0.254$ & this paper \\ 
  Secondary mass [$M_{\odot}$]   & $\approx 0.32$ & this paper  \\
  Secondary radius [$R_{\odot}$] &      $\approx 0.32$  & this paper \\   
  Secondary spectral type &     $\sim$M4V  & this paper  \\     
  \hline\noalign{\smallskip}
\end{tabular}\\
\end{table}
\end{center}

\subsection{Orbital solution}\label{Sect_orbit}

We fit the RV curve of the primary with a Keplerian orbit in the following form \citep[see, e.g.,][]{2020AJ....160..273L}: 
\begin{equation}
    {\varv}_{\rm r} = \gamma + K \left[ \cos{(\nu + \omega)} + e \cos{\omega} \right],
\end{equation}
where $\gamma$ is the mean RV of the system, $K$ is the RV semi-amplitude, $\nu$ is the true anomaly, $\omega$ is the argument of pericenter, $e$ is the eccentricity. The mean anomaly $M$ can be written as
\begin{equation}
    M = \frac{2 \pi}{P_{\rm orb}} (t - t_0) + M_0 = E - e \sin{E},
\end{equation}
where $P_{\rm orb}$ is the orbital period, $M_0$ is the mean anomaly at the reference epoch $t_0$, and $E$ is the eccentric anomaly. The value of $t_0$ was set to the median of RV observation times ($t_0 = 2460927.47244$). Finally, $E$ can be directly converted to true anomaly:
\begin{equation}
    \tan{\frac{\nu}{2}} = \sqrt{\frac{1 + e}{1 - e}} \tan{\frac{E}{2}}.
\end{equation}

We use the affine-invariant Markov chain Monte Carlo (MCMC) ensemble sampler \texttt{emcee} \citep{2013PASP..125..306F} to sample from the posterior with 64 walkers and 100\,000 steps. We introduce two jitter terms for the two different RV sources $\sigma_\mathrm{Konkoly}$ and $\sigma_\mathrm{STELLA}$ to handle underestimated RV uncertainties. Therefore we assume a diagonal covariance matrix in the following form: $\sigma^2_{\mathrm{obs},i} + \sigma^2_{\rm Konkoly/STELLA}$, where $\sigma_{{\rm obs},i}$ is the per-measurement error. As the orbit is close to circular, we sample $h=\sqrt{e} \cos{\omega}$ and $k=\sqrt{e} \sin{\omega}$ instead of $e$ and $\omega$. To make the sampling more stable, we use $\log P_{\rm orb}$ instead of $P_{\rm orb}$. For priors, we use $h^2 + k^2 < 1$ and wide uniform distributions for the other parameters ($\gamma$, $\log P_{\rm orb}$, $K$, $M_0$, $\sigma_{\rm Konkoly}$, $\sigma_{\rm STELLA}$).

\begin{figure}[h]
    \centering
    \includegraphics[width=\columnwidth]{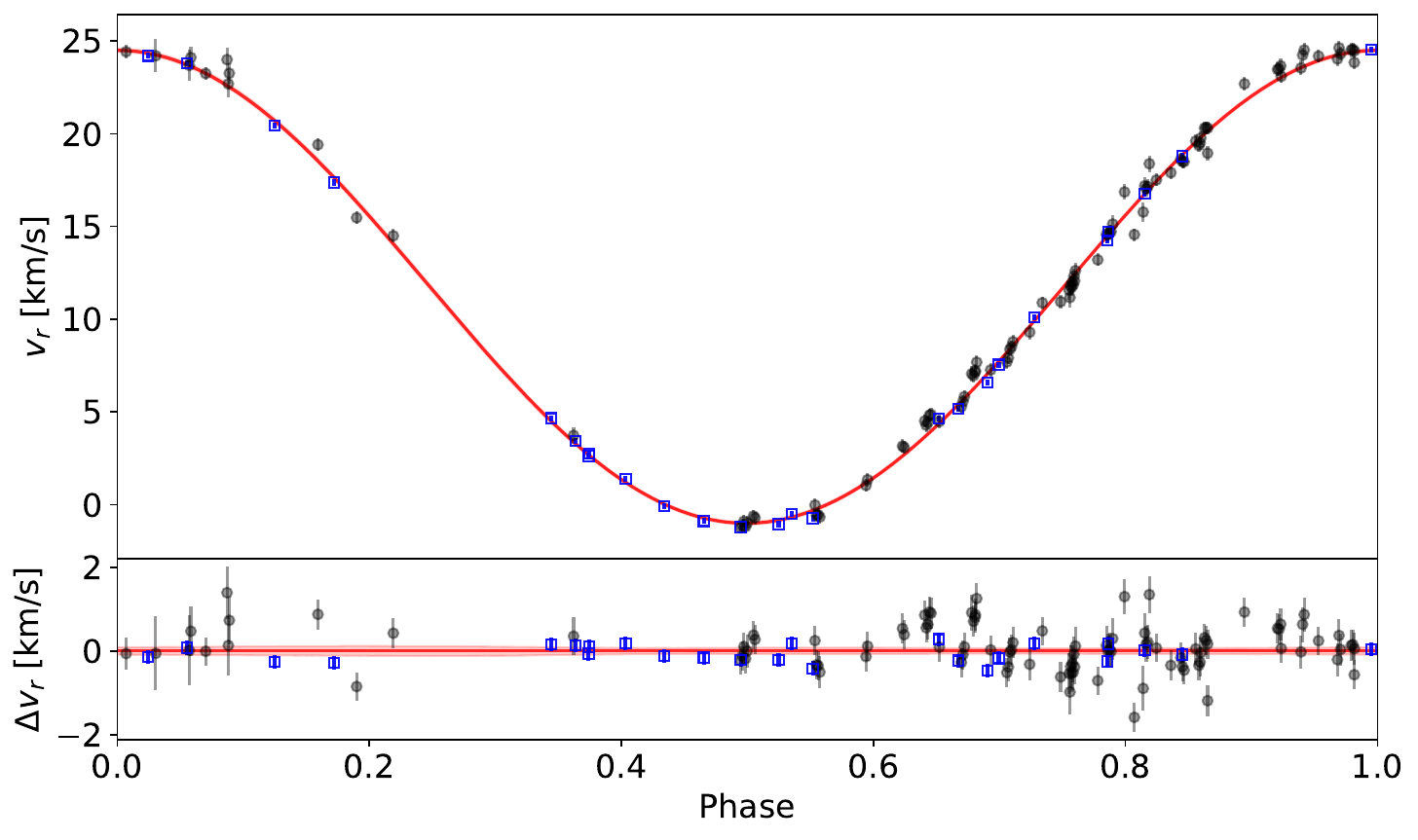}
    \caption{The result of fitting the radial velocity measurements. The black dots are from Piszkéstető Mountain Station of Konkoly Observatory, while the blue open squares are the STELLA data from Iza\~na Observatory. The zero phase has been aligned with the maximum of the RV curve. See also Table\,\ref{tab:RV_params}.}
    \label{fig:RV_fit}
\end{figure}

The result of the RV fit can be seen in Fig.\,\ref{fig:RV_fit}, and the parameters are summarized in Table\,\ref{tab:RV_params}. The probability distribution of the orbital parameters are plotted in Fig.\,\ref{fig:RV_posteior} in Appendix\,\ref{app_posteriors}. The Gelman--Rubin statistic satisfies $\hat{R}<1.01$ for all parameters, indicating good chain mixing and convergence. The posterior is unimodal, with the poorly constrained $M_0$ wrapping around the $2\pi$ boundary. The eccentricity estimated from noisy measurements is known to be biased towards higher values \citep[e.g.,][]{2010ApJ...725.2166H}, so the resulting $e = 0.006_{-0.004}^{+0.005}$ is consistent with an essentially circular orbit. Since we only have the RV curve of the primary, we cannot determine the mass of each component ($M_1$ and $M_2$), only the $f(m)$ mass function:
\begin{equation}
    f(m) = \frac{M_2^3 \sin^3{i}}{(M_1 + M_2)^2} = \frac{P_{\rm orb} K^3}{2 \pi G} (1 - e^2)^{3/2} = (0.0073\pm 0.0001)\,M_\odot.
\end{equation}

Based on the RV fit, the system's ephemeris is:
\begin{equation}
    \mathrm{HJD}_{\phi=0} = 2460928.142_{-0.049}^{+0.050} + 34\fd182_{-0.001}^{+0.001} \times E.
\end{equation}
However, since we have a non-eclipsing binary, following tradition, we assign the zero phase to the maximum of the RV curve, that is, 0.25 phase earlier\footnote{For $e$=0.006 eccentricity, the phase separation between conjunction and the RV maximum deviates from the nominal quarter-cycle characteristic of a fully circularized orbit by only $\sim$0.002 in phase, that is, essentially negligible.} than the primary conjunction. Accordingly, the zero epoch corresponds to the zero phase closest to the median calculated from all available RV measurement times.

\begin{table}[h]
\begin{center}
\caption{Orbital parameters derived from the RV curve fitting}
\begin{tabular}{lr}
\hline\hline
Parameter & Value \\
\hline\smallskip
$P_{\rm orb}$ [days] & $34.182_{-0.001}^{+0.001}$ \\\smallskip
$e$ & $0.006_{-0.004}^{+0.005}$ \\\smallskip
$\gamma$ [km\,s$^{-1}$] & $11.69_{-0.04}^{+0.04}$ \\\smallskip
$K$ [km\,s$^{-1}$] & $12.75_{-0.05}^{+0.05}$ \\\smallskip
$\sigma_{\rm Konkoly}$ [km\,s$^{-1}$] & $0.16_{-0.08}^{+0.06}$ \\\smallskip
$\sigma_\mathrm{STELLA}$ [km\,s$^{-1}$] & $0.16_{-0.03}^{+0.03}$ \\
\hline
\end{tabular}
\label{tab:RV_params}
\end{center}
\end{table}

\subsection{Rotational period}

\begin{figure}[h]
    \centering
    \includegraphics[width=\columnwidth]{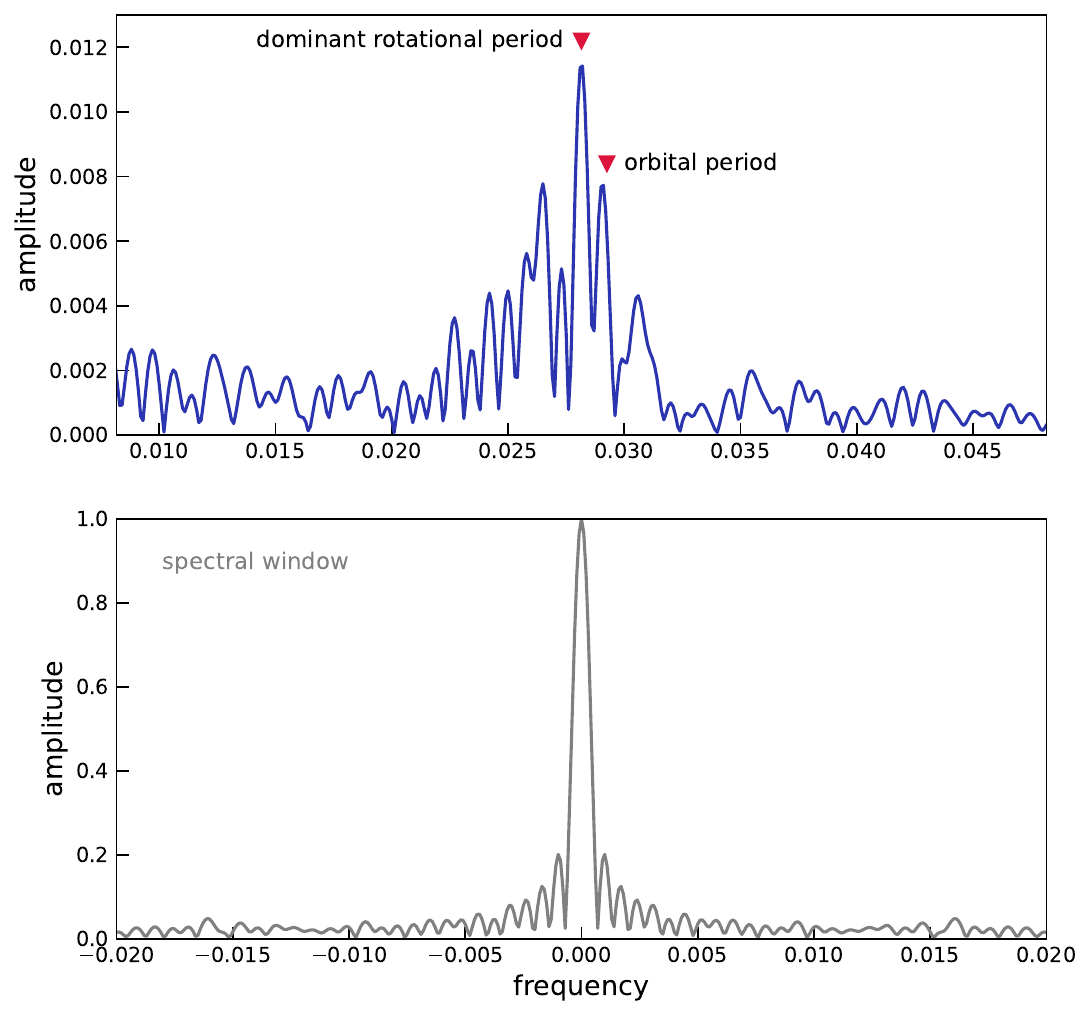}
    \caption{The top panel shows the Fourier amplitude spectrum of KIC\,2852961 based on the full \emph{Kepler} data. The main rotation period and orbital period are indicated by red triangles. The bottom panel shows the spectral window for comparison.}
    \label{fig:rotper}
\end{figure}

To determine the rotation period, we applied the Multi-Frequency Analyzer code \texttt{MuFrAn} \citep{2004ESASP.559..396C} to the entire \emph{Kepler} light curve, which was previously freed from flares to eliminate their potential distorting effect on the result. This Fourier-transformation based frequency analyzer code is suitable for searching for multiple periods in time series of variable stars and displaying the results graphically. The resulting Fourier-spectrum is shown in Fig.\,\ref{fig:rotper}, where the most dominant amplitude occurs at $P_{\rm rot}=35.515^{+0.454}_{-0.226}$ days, which is considered the grand average rotation period caused by the surface spots. This period, however, differs significantly from the orbital period of $P_{\rm orb}=34.182\pm0.001$ days also shown in the periodogram. On the other hand, the absence of a detectable photometric signal at half the orbital period argues against ellipsoidal variability as the origin of the peak at $P_{\rm orb}$. Instead, the presence of a persistent signal exactly at the orbital period is more naturally explained by an active longitude that is phase-locked to the binary orbit by tidal interaction. Such a configuration can coexist with the strongest signal at $P_{\rm rot}=35.515$\,d, which likely traces a dominant spotted region in a certain differentially rotating latitude zone. The amplitude spectrum shows another significant peak at $\sim$37.5\,d, which can also be interpreted as an additional sign of differential rotation.

\begin{figure}[h]
    \centering
    \includegraphics[width=\columnwidth]{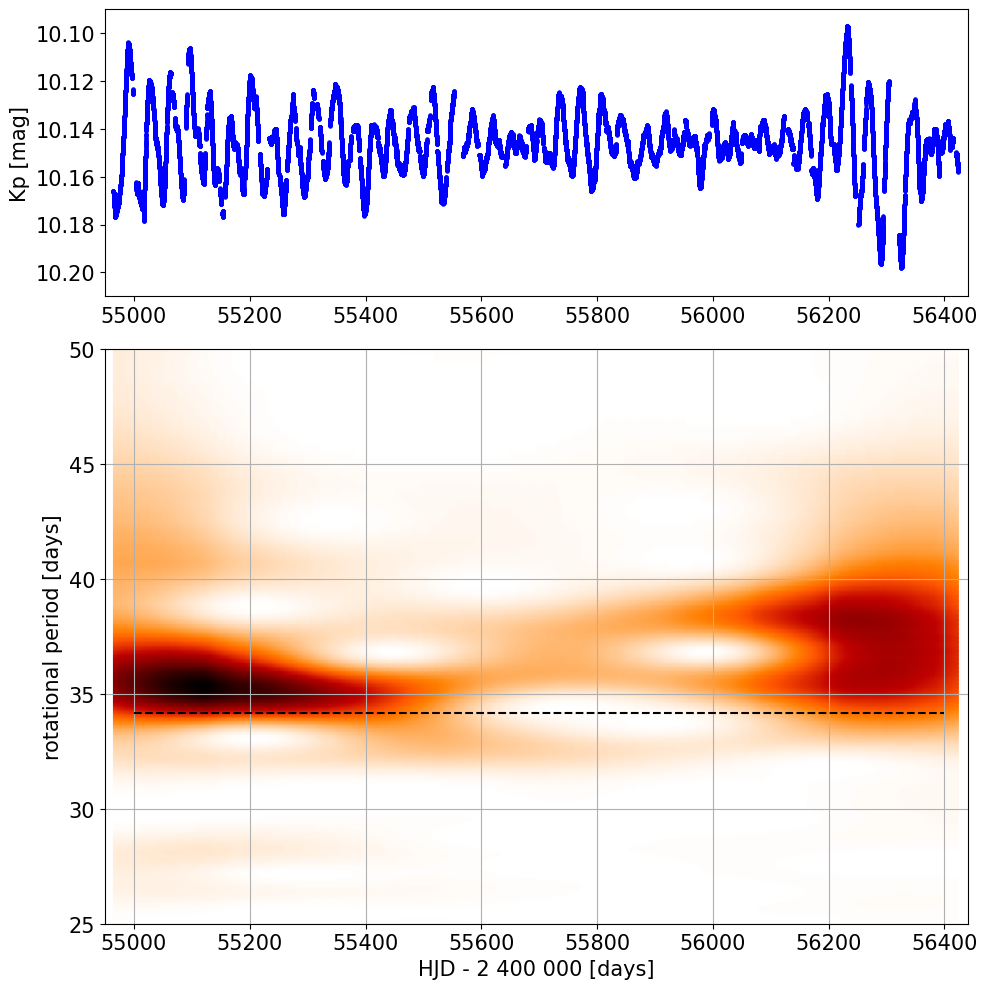}
    \caption{The \emph{Kepler} light curve of KIC\,2852961 with flares removed (top), and the time-frequency diagram \citep{2024JOSS....9.7118B} of the rotation signals (bottom). The  dashed horizontal line on the time-frequency plot marks the orbital period $P_{\rm orb}$=34.182\,d.}
    \label{fig:rotper_stft}
\end{figure}

Figure\,\ref{fig:rotper_stft} shows the temporal evolution of the dominant rotational period signals derived from the \emph{Kepler} light curve of KIC\,2852961 using the windowed Lomb-Scargle periodogram tool from the time-frequency analysis package \texttt{seismolab} \citep[][]{2024JOSS....9.7118B}. Throughout the dataset, the strongest signal remains concentrated near the orbital period, confirming that the giant primary is in a near-synchronous rotational state. However, the distribution of power is not stationary and exhibits a gradual drift toward longer periods, reaching values of $\sim$38\,d during the later part of the \emph{Kepler} observations. In addition, multiple closely spaced period components are present simultaneously, indicating that different active regions rotate with slightly different angular velocities. Such behaviour is difficult to reconcile with rigid-body rotation and is consistent with weak solar-like surface differential rotation, where spots at different latitudes trace different rotation rates. The tendency for the dominant signal to migrate toward periods slightly longer than the orbital period supports the interpretation that the equatorial regions are nearly synchronized by tidal forces, while higher-latitude active regions rotate more slowly. The observed period spread of several days is also compatible with the spot migration inferred from the light-curve modelling and suggests that differential rotation remains detectable despite the strong tidal interaction and nearly circular orbit of the binary system. Further discussion of the indications for differential rotation will be continued in Sect\,\ref{Sect_disc}.

\subsection{Inclination}
Assuming synchronous rotation ($P_{\rm rot}=P_{\rm orb}$), the $R_1$ radius and the $v\sin i$ equatorial velocity--taking into account the estimated uncertainties--yield a formal inclination of 77.7$^{\circ}$$^{+12.3}_{-8.4}$, that is, allowing inclinations from 90$^{\circ}$ down to 69.3$^{\circ}$. On the other hand, the upper limit for the orbital inclination can be given based on the fact that there is no photometric sign of eclipses. Adopting the separation of $a\approx 54.6\,R_{\odot}$ from the orbital solution and assuming a faint red dwarf companion with a radius of $R_2\sim 0.32R_{\odot}$, the non-detection of eclipses implies $i\lesssim 77^{\circ}$. While the nominal inclination of 77.7$^{\circ}$ lies very close to the upper limit imposed by the absence of eclipses, it remains consistent with this constraint within the uncertainties.

Since our close binary system already exhibits a very low orbital eccentricity, indicating advanced tidal evolution (see also Sect.\,\ref{Sect_tidalcirc} below), it is therefore reasonable to assume that substantial spin–orbit alignment has already occurred. Therefore, this upper limit of the orbital inclination can also be considered an upper limit of the rotation axis of the red giant.

\subsection{Primary and secondary masses}

   \begin{figure}[t]
   \centering
   \includegraphics[trim={0.5cm 0cm 1.1cm 1.3cm},clip,width=\columnwidth]{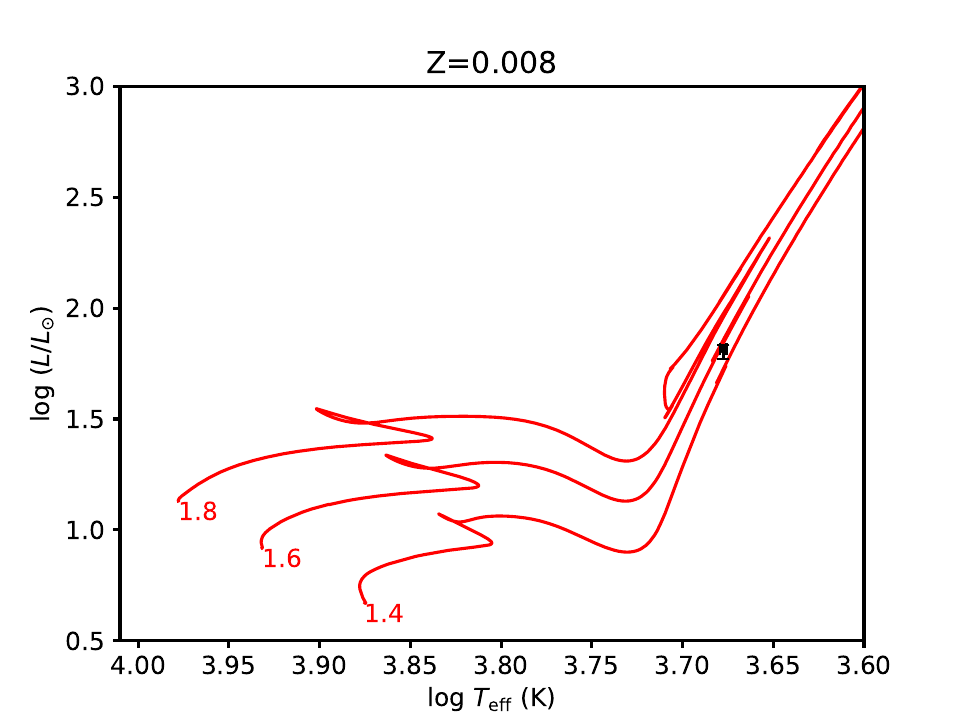}
   \caption{Position of the giant primary of KIC\,2852961 in the H-R diagram (dot). The stellar evolutionary tracks plotted for 1.4, 1.6 and 1.8 solar masses are taken from {\it PARSEC} \citep{2012MNRAS.427..127B}, with $Z$ set to 0.008. From the location, the expected mass is around 1.54\,$M_{\odot}$.}
              \label{fig:hrd}
    \end{figure}

The $T_{\rm eff}$ surface temperature and the luminosity are used to plot the primary star on the Hertzsprung--Russell (H-R) diagram in Fig.\,\ref{fig:hrd}. Stellar evolution tracks are taken from Padova and Trieste Stellar Evolution Code ({\it PARSEC}, \citealt[][]{2012MNRAS.427..127B,2025A&A...701A.258N}) for $Z$=0.008 ([M/H]$\approx$[Fe/H]=$-$0.28). From Fig.\,\ref{fig:hrd}, we estimate a stellar mass of 1.54$\pm 0.2$\,$M_{\odot}$ for the K giant with an age of $\approx$1.5\,Gyr (i.e. around the red giant bump), which is only a slight decrease in mass compared to \citetalias{2020A&A...641A..83K}. The mass and the radius of the giant would yield a surface gravity of $\log g$=2.46$\pm$0.06, in agreement with the value obtained from spectral synthesis and listed in Table\,\ref{tab1}. 

The mass of the unseen secondary is derived from the spectroscopic mass function $f(m)$, substituting the primary mass inferred from the H-R diagram and adopting $i$=70$^{\circ}$ as a representative value of the orbital inclination, also consistent with the non-detection of eclipses. All this yields a companion mass of $M_2\approx 0.32\,M_{\odot}$, corresponding to a late-type main-sequence M dwarf of approximately spectral type M4V. Based on empirical mass–radius relations for low-mass main-sequence stars, the expected radius of the secondary is $R_2\approx 0.32\,R_{\odot}$, sufficiently small to explain the absence of detectable spectral trace. Applying the Stefan$-$Boltzmann relation and using the data above, we obtain a luminosity ratio of $L_2/L_1 \approx 1.6 \times 10^{-4}$; thus, the companion star's contribution to the total luminosity can indeed be considered negligible.

\section{Activity signatures from space photometry}\label{Sect_metandres}

\subsection{Flare-frequency distribution}\label{Sect_FFD}

\begin{figure}[t]
   \centering   \includegraphics[width=\columnwidth]{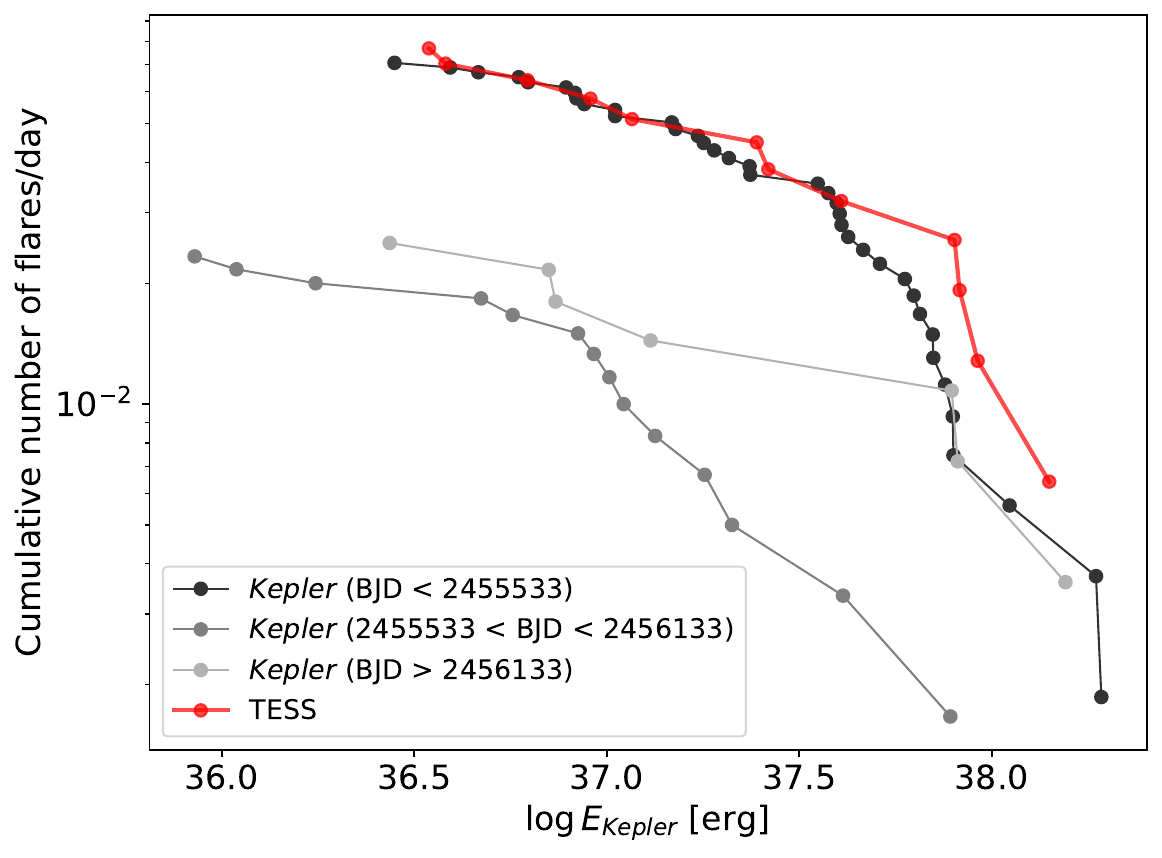}
   \caption{Comparison of cumulative flare-frequency distributions derived from \emph{Kepler} data (gray scale) and TESS data (red). The \emph{Kepler} flares are taken from \citetalias{2020A&A...641A..83K}, but their energies are recalculated with using the corrected quiescent \emph{Kepler} luminosity of the star. To ensure comparability, TESS flare energies are converted to the \emph{Kepler} band as described in Sect.\,\ref{Sect_FFD}. The \emph{Kepler} dataset is divided into the same three segments as in Fig.\,13 of \citetalias{2020A&A...641A..83K}: black dots correspond to the first segment, when the rotational amplitudes were $\Delta F_{\rm rot}\sim3$-7\%; the FFD plotted in gray represents the small amplitude interim period ($\Delta F_{\rm rot}\sim2$\% on average); the third FFD in light gray corresponds to the third part of the \emph{Kepler} light curve, when the rotational amplitude increased again from $\Delta F_{\rm rot}\sim3$\% up to $\sim$7\%.}
   \label{fig:ffd}
\end{figure}

We detected a total of 12 flares in the six available sectors of TESS data, 11 new ones plus one already reported in \citetalias{2020A&A...641A..83K}. The flare energies in the TESS bandpass were determined following the method in \citetalias{2020A&A...641A..83K}, with using $7.21\times10^{34}$\,erg\,s$^{-1}$ for the quiescent TESS-band luminosity, after correcting for the newly determined $T_{\rm eff}$ and stellar radius. The flare parameters are summarized in Table\,\ref{table:TESSflares}. Figure\,\ref{fig:ffd} shows the flare frequency distribution from TESS, compared to the \emph{Kepler} data taken from \citetalias{2020A&A...641A..83K}, also corrected for the newly derived quiescent \emph{Kepler} luminosity of $6.59\times10^{34}$\,erg\,s$^{-1}$. We note that the correction yielded a non-significant decrease of $\sim$8\% in flare energy; consequently, there is essentially no change to the energy-duration diagram in \citetalias{2020A&A...641A..83K} either \citep[cf. also][their Fig.\,14]{2021PASJ...73...44M}. However, the energies of TESS and \emph{Kepler} flares are not directly comparable due to the different spectral response functions. Therefore, assuming a characteristic flare temperature of $\sim$9000\,K, in Fig.\,\ref{fig:ffd} the TESS flare energies were converted to the \emph{Kepler} band according to \citet[][see their Eq.\,10]{2021A&A...650A.138S}. The comparison of the TESS and \emph{Kepler} FFDs should nevertheless be interpreted with some caution because only 12 TESS flares were detected and the observing cadence varies among the sectors. Nine of the detected flares were observed with a 600-s cadence, while two were detected in Sector 14 with a 1800-s cadence and one in Sector 74 with a 200-s cadence. The flare durations, ranging from 6.0\,h to 41.5\,h, are significantly longer than even the 1800-s sampling interval. Thus, the different cadences are not expected to cause significant temporal smearing of the detected events. Differences in sampling and photometric sensitivity may nevertheless affect the completeness of the flare sample, particularly at the low-energy end. We therefore regard the TESS FFD primarily as an independent consistency check of the strong flare activity identified from the \emph{Kepler} data.

Unfortunately, a full rotation period does not fit into a TESS sector, so we cannot measure accurate rotation amplitudes for each sector. Although the available consecutive TESS sectors (i.e., Sectors 40-41 and 54-55) span more than one rotation cycle, the individual sectors require separate normalization and detrending, which introduces some uncertainty regarding the amplitude of the rotational modulation across sectors. Furthermore, the temporal separation of the available TESS sectors does not allow us to determine how persistent a given rotation amplitude may be. Anyhow, the rough estimates of the rotational amplitudes for the six TESS sectors fall between \mbox{$\Delta F_{\rm TESS} \sim 3$--4.5\%} (see Fig.\,\ref{fig:TESS_lc} in Appendix\,\ref{sect:appendix_TESS_lc}) which, however, need to be converted to \emph{Kepler} filter before comparison. The observed amplitude can be written approximately as
\begin{equation}
\Delta F_{\rm rot} \approx f_{\rm spot}\left(1-\frac{F_{\rm spot}}{F_{\rm phot}}\right),\label{eqAmp}
\end{equation}
where $f_{\rm spot}$ is the spot filling factor and $F_{\rm spot}$ and $F_{\rm phot}$ are the band-integrated surface fluxes for the spotted and the unspotted area, respectively. The ratio of \emph{Kepler} and TESS amplitudes is therefore
\begin{equation}
\frac{\Delta F_{\rm Kep}}{\Delta F_{\rm TESS}} = \frac{1-{F_{\rm spot}^{\rm Kep}}/{F_{\rm phot}^{\rm Kep}}}{1-{F_{\rm spot}^{\rm TESS}}/{F_{\rm phot}^{\rm TESS}}}.\label{eqAmpratio}
\end{equation}
Assuming $T_{\rm spot}$$\approx$3800\,K with $T_{\rm phot}$=4760\,K and using blackbody approximation through the two filters, the ratio of $\Delta F_{\rm Kep}/{\Delta F_{\rm TESS}}$ is $\approx$1.15. That is, amplitudes of 3--4.5\% in TESS would appear slightly larger, $\approx$3.5--5.2\% in the \emph{Kepler} band. These values are more similar to the large amplitude \emph{Kepler} observations. Accordingly, the FFD shown in Fig.\,\ref{fig:ffd} for TESS flares (plotted in red) best overlaps with the FFD representing the large amplitude \emph{Kepler} period with the most frequent and most energetic flares. This finding provides additional support for the connection already suggested in \citetalias{2020A&A...641A..83K} between large spot-induced rotational amplitudes and frequent and high-energy flares.

\begin{center}
\begin{table}
\centering
\caption{TESS Sector number, flare peak time, duration, equivalent duration, and integrated flare energy in the TESS bandpass for the 12 flares detected in the available TESS light curves.}\label{table:TESSflares}
\begin{tabular}{c r r r r}
\hline\hline\noalign{\smallskip}
TESS & $T_{\rm peak}$  & $\Delta t$ & $t_{\rm eq,TESS}$ & $E_{\rm f,TESS}$ \\
Sector & BJD--2458000 & [h]  & [sec]  & [$10^{36}$\,erg] \\
\hline\noalign{\smallskip}
14 & 686.301 & 21.0 & 64 & 4.6 \\
14 & 695.468 & 30.0 & 948 & 68.4 \\
40 & 1397.805 & 41.5 & 271 & 19.5 \\
40 & 1404.069 & 15.8 & 1454 & 104.9 \\
41 & 1421.791 & 30.8 & 419 & 30.3 \\
41 & 1423.972 & 17.3 & 120 & 8.7 \\
41 & 1436.659 & 13.2 & 94 & 6.7 \\
41 & 1441.868 & 6.0 & 39 & 2.8 \\
54 & 1785.368 & 33.0 & 851 & 61.4 \\
55 & 1807.034 & 12.8 & 36 & 2.6 \\
55 & 1816.617 & 39.8 & 825 & 59.6 \\
74 & 2333.602 & 19.1 & 253 & 18.2 \\
\hline\noalign{\smallskip} 
\end{tabular}\\
\end{table}
\end{center}

\subsection{Time-series spot modelling}\label{Sect_SML}

For spot modeling, we use the full long-cadence \emph{Kepler} light curve, from which flares were previously removed. We use two different codes, thus allowing for a more thorough evaluation of the results. One is SML based on the analytic equations of \citet{1977Ap&SS..48..207B}; for a more detailed description, see \citet{2002KOTN...12....1R} and \citet{2003AN....324..202R}. The other is the open-source code SpotModel\footnote{\url{https://github.com/blanka12345/SpotModel}} Python package, which implements the analytical starspot model of \citet{1987ApJ...320..756D}. Both the SML and SpotModel codes represent the stellar surface with a small number of homogeneous circular spots and computes synthetic light curves including linear limb darkening. The spot longitudes ($\lambda_i$), latitudes ($\beta_i$), angular radii ($\gamma_i$), and temperature contrasts ($\Delta T_i$) are optimized to reproduce the observed photometric variations. By modelling consecutive light-curve segments, SpotModel, just like the SML code, enables time-series spot modelling, allowing the temporal evolution of the spot configuration to be followed through changes in the spot parameters.

The inversion of photometric light curves is inherently non-unique \citep[e.g.][]{2021AJ....162..123L}, so we restricted ourselves to a simple two-spot model, keeping the number of free parameters low \citep{1997A&A...323..801K}. With the spot temperatures fixed at $-$1000\,K relative to the 4760\,K surface (assumed "unspotted"), this ultimately results in six free parameters: the astrographic coordinates of the centres of the spots ($\phi_1$, $\beta_1$, $\phi_2$, $\beta_2$) and the spot radii ($\gamma_1$, $\gamma_2$). This usually allows the inversion code to provide robust constraints on the phase distribution and temporal evolution of the dominant active regions. We note here that the uncertainty arising from not knowing the true unspotted brightness significantly affects the resulting spot coverage. Therefore, the spot coverage (spot filling factor) derived from our spot models cannot be considered an absolute parameter when compared with other stars.

\begin{figure*}[t]
   \centering
   \includegraphics[width=1.9\columnwidth]{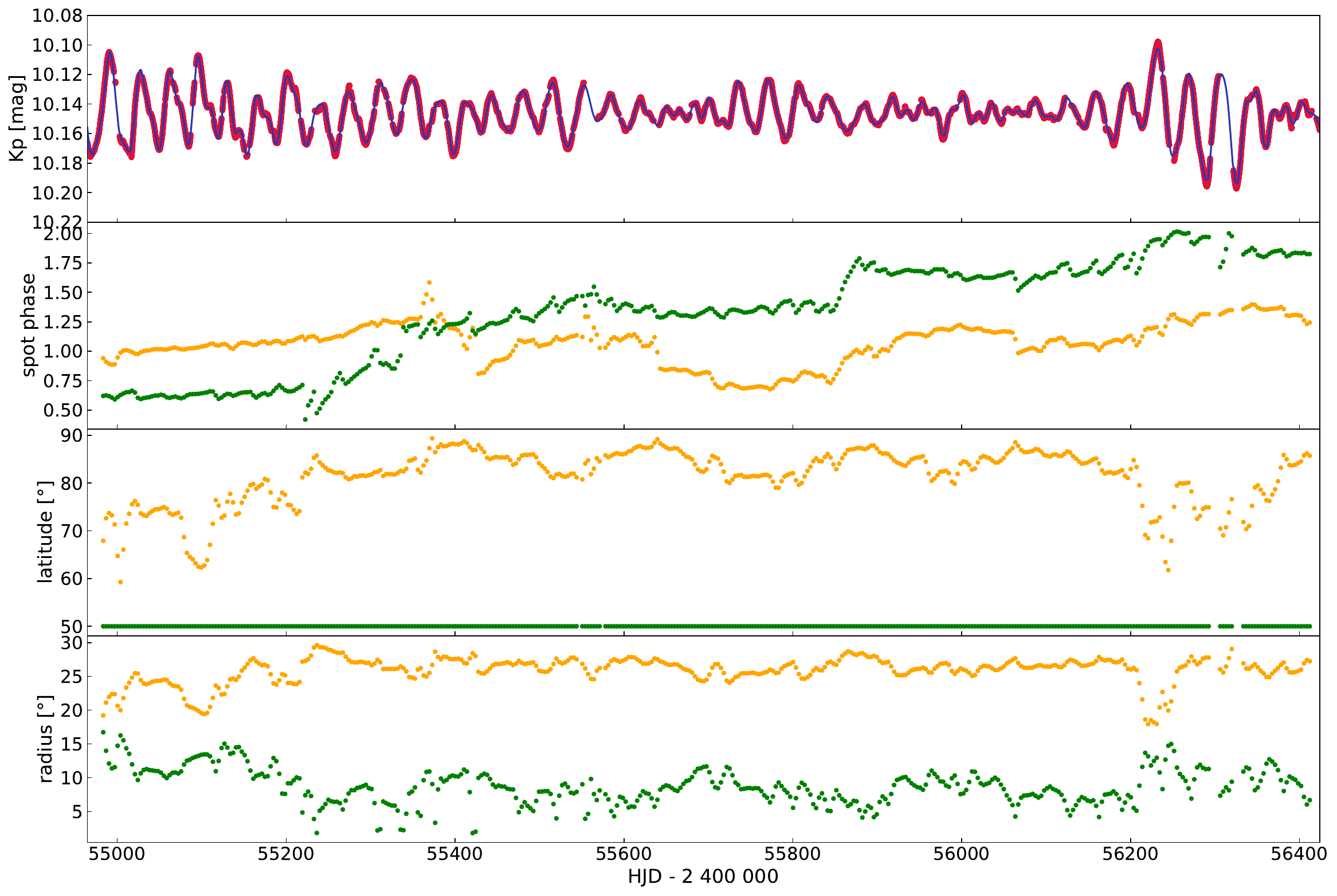}
   \caption{The result of our two-spot model using SML, in which the latitude coordinate of one spot was fixed at $\beta$=50$^{\circ}$. The top panel shows the time-series model fit (blue) of the full \emph{Kepler} light curve (red), while the panels below show the current spot coordinates (spot phases, latitudes and radii). The coordinates of the two spots are represented in different colors (orange and green).}
   \label{fig:SML}
\end{figure*}

The first attempt at fitting the light curve with six free parameters (i.e., two times three free coordinates of the two spots) is shown in Fig.\,\ref{fig:attempt1_lc} in Appendix\,\ref{sect:appendix_SML}, while the change in the corresponding spot coordinates is shown in Figs.\,\ref{fig:attempt1_spotpars}-\ref{fig:attempt1_poleon} in Appendix\,\ref{sect:appendix_SML}. Although the fit to the highly variable light curve is demonstrated to be satisfactory (see Fig.\,\ref{fig:attempt1_lc}), this first experiment led to difficult-to-interpret results for both SpotModel and SML, that is, for example, hemispheric degeneracy due to the high inclination, often overlapping spots, multiple swapping of the latitude coordinates of the two spots simultaneously with the radii, all of which made it impossible to track the evolution of individual spots. Therefore, in our second experiment, as an additional constraint, we fixed the latitude coordinate of one of the spots at 50$^{\circ}$, because based on the initial results it seemed that the less dominant spot appeared mostly around this latitude. The resulting five-free-parameter spot model using SML is plotted in Fig.\,\ref{fig:SML}, which shows the general picture that the high latitude spot covering the visible pole is responsible for adjusting the highly fluctuating brightness level, while the other, less dominant spot, which shows significant phase drift over the entire time interval of the \emph{Kepler} light curve, can be attributed to the differential rotation signal (see further discussion in Sect.\,\ref{sect_disc_dr}). For comparison, the results of the five-free-parameter fit using SpotModel is plotted in Fig.\,\ref{fig:attempt2_SpotModel} in Appendix\,\ref{sect:appendix_SML}. We note, that the similarity is clear in terms of the resulting spot parameters, which demonstrates the consistency between the two implementations under the adopted spot-model assumptions. However, in the case of SML, we see slightly smoother transitions in the spot phases where either datapoints are missing or the shape of the light curve changes abruptly.

\subsection{Spot positions and flare occurrences}

Several studies have investigated the possible relationship between starspots and flares, two key manifestations of stellar magnetic activity, and the nature of this relationship \citep[see, e.g.,][]{2014ApJ...797..121H,2018ApJ...868....3R,2023ApJ...948...64I}. One possible approach is to associate each flare with the nearest starspot in rotational phase and compare the resulting phase difference distribution with that obtained from randomly placed points (i.e., equivalents to "random flares") in the light curve \citep[][see their Sect.\,3.5]{2026A&A...706A..54G}. This comparison can then be evaluated, for example, with the Kolmogorov--Smirnov test \citep[see, e.g.,][]{ivezic2020statistics}. If flares are linked to active regions, they should occur preferentially when the corresponding spot is visible, resulting in similar flare phases and spot longitudes.

For the test, we took the spot phases from the six-free-parameter spot model, but we emphasize that the spot phases are the most robust parameters, which is why they show a high degree of agreement when comparing our six- and five-free-parameter models (cf. the spot phases in the corresponding panels of Fig.\,\ref{fig:attempt1_spotpars} and Fig.\,\ref{fig:SML}). When comparing the phase difference distribution of $10^6$ randomly placed flares on the Kepler light curve with the phase difference distribution of the 59 real flares, the Kolmogorov--Smirnov test yielded a $p$-value of 0.48, thus it failed to reject the null hypothesis that the two datasets come from the same parent distribution. A significantly lower $p$-value, such as $p < 0.001$, would indicate a possible correlation between flare phases and starspot longitudes. However, our results show no evidence of a correlation between starspot longitudes and flare phases, suggesting that flare phases mostly follow a random distribution, consistent with the recent result of \citet[][]{zhang2025starspots}. Nevertheless, this finding does not necessarily mean that spots and flares are not related; it is possible that there are several smaller active regions where flares originate, while the naturally simplified spot model fits another, more dominant active region, or it is also possible that flares can be linked to a single active region located at high latitudes, which remains visible for most of the rotation period \citep{2014ApJ...797..121H}. For more recent discussions on this issue, see, e.g., \citet{2021MNRAS.507.1723I,2024A&A...682A.176B}.

\subsection{Flare occurrence along the orbital phase}

\begin{figure}[thb]
    \centering
    \includegraphics[angle=-180,width=\columnwidth]{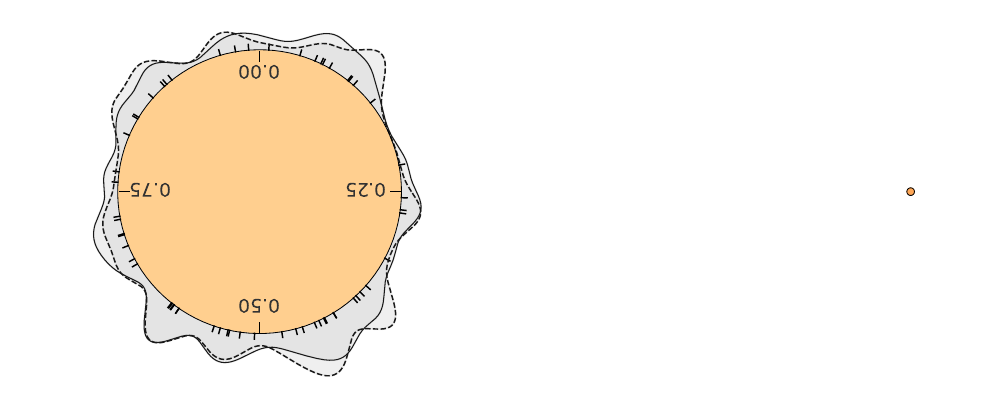}
    \caption{Phase distribution of flares in the pole-on view of the system. The orbital motion is counterclockwise, and the system is shown at zero orbital phase. (The observer's direction is therefore at six o'clock at this moment.) The two circles represent the two components with radii and separation to scale. 
     Ticks around the giant primary mark the phases of flares in the orbital reference frame, and the gray zones with solid and dashed contours show Gaussian KDEs of the phase distribution unweighted and weighted with flare energies, respectively.}
    \label{fig:flare_phases}
\end{figure}

In \citetalias{2020A&A...641A..83K} (see its Fig.\,14) we argued that the distribution of flares according to the rotational phase is highly likely to be random. Since we did not yet know the binary orbit, we used an arbitrary phase, and instead of $P_{\rm orb}$, we calculated with the approximate photometric period ($P_{\rm phot}$=35.5\,d) fitted to the rotational modulation. Now that we have a well-defined orbital period of 34.182\,d and the zero phase is matched to the maximum of the RV curve, we repeat the phase ordering of the flares.

The distribution of flare phases from our \emph{Kepler} dataset in the reference frame of the binary orbit is shown in Fig.\,\ref{fig:flare_phases}. Solid and dashed lines show Gaussian Kernel Density Estimates (KDEs) of the phase distribution unweighted and weighted with flare energies, respectively. To assess whether flare phases show any significant preference in the orbital reference frame, we employed the Rayleigh test \citep{circ_stat} using \texttt{rayleightest} from \texttt{astropy} \citep{astropy:2022}. We tested for the flare phases with and without weighting with flare energy. The test could not reject the null hypothesis of uniform phase distribution with $p$=0.23 for the unweighted case, and $p$=0.17 in the energy weighted case. As the Rayleigh test is only sensitive to unimodal distributions, we also employed the Kuiper test \citep{kuiper1960tests} for the unweighted dataset using the \texttt{astropy} implementation. The Kuiper test also failed to reject the null hypothesis of uniform phase distribution with $p$=0.20. That is, we find no statistically significant departure from a uniform phase distribution in the \emph{Kepler} flare sample. Finally, we note that the near-uniform \emph{Kepler} exposure coverage over orbital phase is statistically consistent with uniform sampling, indicating that the phase-folded photometric signals are free from significant systematic sampling bias.

\section{Discussion}\label{Sect_disc}

\subsection{A late-stage tidally circularizing system}\label{Sect_tidalcirc}

According to the orbital data in Sect.\,\ref{Sect_orbit}, the binary system exhibits a very low eccentricity of $e$=0.006, an almost full tidal circularization, indicating an advanced stage of tidal evolution. In the framework of equilibrium tide theory \citep{1977A&A....57..383Z,1989A&A...220..112Z} the circularization time $\tau_{\rm circ}$ scales steeply with the ratio of orbital separation to stellar radius, $\tau_{\rm circ}\propto [q(1+q)]^{-1}\left(\frac{a}{R_1}\right)^8$, implying a dramatic reduction in $\tau_{\rm circ}$ as the primary star expands along the giant branch. That is, in the case of KIC\,2852961, the efficiency of tidal circularization is expected to be very high, which is due to the increase in radius resulting from the primary's evolutionary phase, which also entails the deepening of the giant's convective envelope.

\subsection{Differential rotation}\label{sect_disc_dr}

The slight but significant difference between the orbital period ($P_{\rm orb}$=34.182\,d) and the dominant photometric rotation period ($P_{\rm rot}$=35.515\,d) may provide evidence for surface differential rotation on the giant primary. A plausible interpretation is that tidal torques have synchronized the equatorial regions of the star to the orbital motion, while the observed spots are located at higher latitudes that rotate more slowly. In this case, the measured photometric period would not represent the equatorial rotation rate but rather the rotation period of the latitudinal zone of the visible surface that is more covered with spots. Assuming $P_{\rm orb}$ approximates the equatorial rotation period, the estimate of relative surface shear is $\alpha_{\rm DR}$$\gtrsim$$\Delta P/P_{\rm orb}$$\sim$0.04, that is, a moderate solar-like differential rotation. The lower limit of 0.04 corresponds to the extreme case where the most spotted zone is furthest from the equator, that is, at the pole--which, however, contradicts the observed large rotation amplitudes. If the most spotted zone is at a lower latitude, e.g. around $\beta$=60$^{\circ}$, $\alpha_{\rm DR}$ would be 0.05, while $\beta$=45$^{\circ}$ spot latitude would yield $\alpha_{\rm DR}$=0.074. Moreover, considering the additional rotational signal appearing in the Fourier spectrum at $\sim$37.5\,d, the estimated relative surface shear may be even stronger. On the other hand, in the case of a small degree of subsynchrony at the equator, the resulting shear would be somewhat reduced. Anyhow, a conservatively estimated shear coefficient of $\alpha_{\rm DR}$$\approx$0.05$\pm$0.02 for KIC\,2852961 based on the \emph{Kepler} data is compatible with values of a few percent long observed for evolved components in other RS\,CVn-type systems \citep[e.g.,][etc.]{1984A&A...135..255B,1985A&A...148...29B,1986A&A...156..106B,1995ApJS...97..513H,2004MNRAS.348.1175P,2004SoPh..224..187A,2007A&A...463.1071K,2012A&A...539A..50K,2014IAUS..302..379K,2016A&A...593A.123O,2020MNRAS.492.3647X}, where large convective envelopes support latitude-dependent rotation while tidal forces act to synchronize the rotation of the star.

In the case of KIC\,2852961, tidal interactions are expected to be effective, given a separation of only $a$$\approx$$54.6R_\odot$, corresponding to $a/R_1$$\approx$4.6. The nearly circular orbit indicates that substantial tidal evolution has already taken place, suggesting that synchronization should also be well advanced. Nevertheless, complete synchronization of a giant with deep convective bulk may not occur uniformly across all latitudes, particularly if differential rotation is maintained by the interplay between convection and rotation. Moreover, the Fourier amplitude spectrum (see Fig.\,\ref{fig:rotper}) exhibits several closely spaced rotational frequencies around the dominant rotational peak. Such frequency structures are commonly interpreted as signatures of spots located at different latitudes and rotating with slightly different angular velocities. 

The result of our spot modelling, although following a two-spot simplification, is quite complex (see Sect\,\ref{Sect_SML}). Nevertheless, our second modelling attempt with a spot fixed at 50$^{\circ}$ latitude shows a relatively clear picture during which the phases of the two spots can be followed well: the higher latitude spot (usually covering the pole) does not change much in phase, while the phase of the spot at 50$^{\circ}$ latitude increases continuously (i.e. lags behind the orbital period). This may even indicate a departure from rigid synchronous rotation, consistent with differential rotation. In Fig.\,\ref{fig:SML} the estimated average retrograde phase shift of the lower latitude spot is $\approx$0.356$^{\circ}$/d relative to the orbital period. Given that the spot is at $\beta$=50$^{\circ}$ latitude (cf. Fig.\,\ref{fig:SML}), and assuming that the equatorial regions are synchronized with the orbit ($P_{\rm eq}$=$P_{\rm orb}$=34.182\,d), the inferred surface shear parameter is $\alpha_{\rm DR}$$\approx$0.058. This value, although model-dependent due to the constraints of spot modeling, is in a good agreement with the estimate above, which was derived from the difference between $P_{\rm orb}$ and the dominant photometric rotation periods. Finally we note that the role of the higher latitude spot (which usually also covers the pole) is limited here, as although it can contribute significantly to the change in the average flux, it contributes only slightly to the rotational modulation.

Although spot emergence, decay, and finite spot lifetimes can also produce complex or time-dependent structures, as shown in the periodograms (see Figs.\,\ref{fig:rotper}-\ref{fig:rotper_stft}), the systematic spot migration found from our time-series spot modelling provides an independent indication consistent with differential rotation. Therefore, it is more reasonable to consider the presence of multiple, close periods as further support for differential rotation.

\begin{figure}[thb]
    \centering
    \includegraphics[trim={1.0cm 0.5cm 0.7cm 0.7cm},clip,width=\columnwidth]{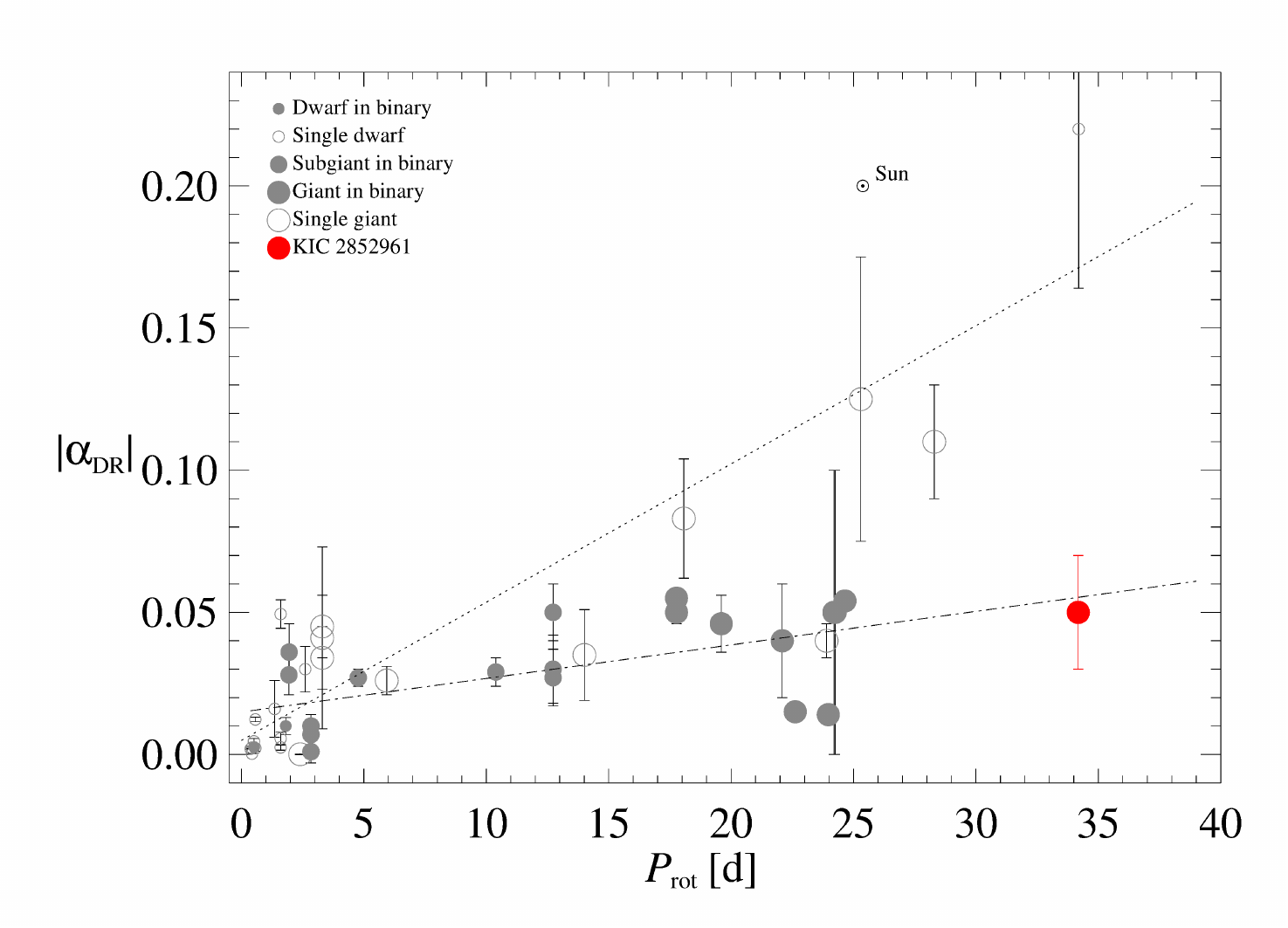}
    \caption{Absolute value of the surface shear coefficient $\alpha_{\rm DR}$ as a function of the rotational period $P_{\rm rot}$ for spotted stars (\citealt{2017AN....338..903K}, last updated in \citealt{2026A&A...706A..54G}). Single stars are indicated by circles, and those in close binary systems are denoted by filled symbols. KIC\,2852961 is marked in red. The dotted line is a linear fit to single stars, and the dash-dotted line with a significantly lower slope is to stars in close binary systems, suggesting the fundamental role of binarity in surface shear confinement.}
    \label{fig:prot-alpha}
\end{figure}

From an empirical perspective, such a value of $\alpha_{\rm DR}$ is fully consistent with expectations for tidally interacting binaries. Based on a compiled sample of stars, \citet{2017AN....338..903K} showed that the rotation–differential rotation relationship of single stars differs significantly from that of stars in close binary systems: while the relative shear of single late-type stars generally increases with rotation period, stars in close binaries exhibit substantially weaker differential rotation, suggesting that tidal forces act as an effective regulator of the latitudinal shear. In their sample, many RS\,CVn-type binaries with evolved components show $\alpha_{\rm DR}$ values of only a few percent, comparable to the value inferred for KIC\,2852961. Based on all this, there is good reason to believe that the primary component of KIC\,2852961 is likely a near-synchronous giant that exhibits a moderate solar-like differential rotation. With this in mind, we plotted the star on the $P_{\rm rot}-\alpha_{\rm DR}$ diagram \citep{2017AN....338..903K}. The resulting Fig.\,\ref{fig:prot-alpha} suggests that tidal evolution has already been highly effective and likely synchronized the equatorial layers of our giant star, while not completely suppressing the differential rotation maintained by the deep convective envelope.

\subsection{Spot modulation amplitudes vs. flare energies and flare frequency}

In the following, we discuss the relationship between the photometric variability due to spots and the resulting flare energy and flare occurrence rate. According to \citet{2013ApJ...771..127N}, the magnetic energy stored in the vicinity of a starspot with characteristic size $l$ and average magnetic field strength $B$ can be estimated based on $E_B\approx \frac{B^2l^3}{8\pi}$, while the amplitude $\Delta F_{\rm rot}$ of the rotational light curve due to spots can be given as

\begin{equation}
\Delta F_{\rm rot} \propto \left[ 1 - \left(\frac{T_{\rm spot}}{T_{\rm phot}}\right)^4 \right] \frac{A_{\rm spot}}{A_{\rm phot}},\label{eqDeltaFrot}
\end{equation}
where $A_{\rm spot}$ $(\sim l^2)$ is the total area of visible starspot, $A_{\rm phot}$ is the projected stellar surface area, and $T_{\rm spot}$ and $T_{\rm phot}$ are the spot and photospheric temperatures, respectively. 
The energy released by a flare from the vicinity of the spot is necessarily less than the total stored energy, that is,
\begin{equation}
E_{\rm flare}^{\rm max}\leq E_B\propto B^2A_{\rm spot}^{3/2}.\label{eqEflare}
\end{equation}
Based on Eq.\,\ref{eqDeltaFrot}-\ref{eqEflare}, it can be simply written that
\begin{equation}
E_{\rm flare}^{\rm max}\leq E_B\propto  \Delta F_{\rm rot}^{3/2}.\label{EmaxfvsDFrot}
\end{equation}
For KIC\,2852961, using the characteristic $E_0 \sim 10^{38}$\,erg flare energy measured in the \emph{Kepler} bandpass together with a typical \emph{Kepler} light curve amplitude (due to spots) of $\Delta F_{\rm rot} \sim 0.03$ provides a predictive relation of
\begin{equation}
E_{\rm flare}^{\rm max} \lesssim E_0 \left(\frac{\Delta F_{\rm rot}}{0.03}\right)^{3/2}.\label{eq8}
\end{equation}
Applying this scaling to different activity states of KIC\,2852961, that is, for periods where the light curve amplitudes due to spots differ significantly, we find that during epochs of enhanced spot modulation ($\Delta F_{\rm rot} \sim 0.06$), the characteristic flare energies in the \emph{Kepler} band increase to $E_{\rm flare}^{\rm max} \sim 3\times10^{38}$\,erg, whereas during lower-activity phases ($\Delta F_{\rm rot} \sim 0.02$), they decrease to $E_{\rm flare}^{\rm max} \sim 5\times10^{37}$\,erg, implying a factor of $\sim$6 increase between the low- and high-activity (i.e., spot-amplitude) states.

The relation between flare occurrence and spot-induced brightness variability can be  considered by combining the flare frequency distribution (FFD) with the dependence of the maximum available flare energy on $\Delta F_{\rm rot}$ suggested by Eq.\,\ref{eq8}. This relation, however, does not by itself imply a corresponding increase in the flare occurrence rate at a fixed energy threshold. Such a dependence would require an additional assumption about how the normalization of the FFD varies with the available magnetic-energy scale. Empirical support for such an activity-dependent FFD normalization is provided by \citet{2017PASJ...69...41M}, who found that the flare occurrence frequency increases with starspot area while the slope of the FFD remains approximately unchanged. Thus, increasing spot coverage appears to raise the overall level of flare activity without necessarily changing its energy dependence. Although the fitted FFD of KIC\,2852961 in \citetalias{2020A&A...641A..83K} is a broken power law, with different slopes below and above the break energy, its normalization can likewise vary independently of these slopes. We therefore interpret the higher flare occurrence rate associated with larger rotational modulation amplitudes as implication for a larger magnetic-energy reservoir, reflected primarily in an enhanced FFD normalization rather than in a change of the power-law slopes. This interpretation is consistent with the fact that larger rotational modulation amplitudes of KIC\,2852961 are associated with an increased (super)flaring rate. This seems to be supported by observations of other flaring stars including the Sun \citep[e.g.,][etc.]{2013ApJ...771..127N, 2013PASJ...65...49S, 2015EP&S...67...59M, 2022ApJ...936...49K, 2025ApJ...985..158T}.

\subsection{X-ray emission}
No targeted X-ray observations are available for KIC\,2852961\footnote{\url{https://heasarc.gsfc.nasa.gov/cgi-bin/W3Browse/w3browse.pl}}. Thus we searched for detections in various all-sky surveys. In the ROSAT All-Sky Survey (RASS), the star was detected with a count rate of $0.0708\pm0.0135$\,cts\,s$^{-1}$ \citep{Boller2016}. Using the count-to-flux conversion for giant stars derived by \citet{Huensch1996} and the star's distance given in Table\,1 results in an X-ray luminosity of $(6.18\pm1.22)\times10^{31}$\,erg\,s$^{-1}$ in the RASS 0.1-2.4\,keV band. The fractional X-ray luminosity amounts to $\log{(L_\mathrm{X}/L_\mathrm{bol})}=-3.62$, comparable to the most active stars in a sample of nearby late-type giants \citep{Huensch1998}, indicating its high level of activity.

The position of KIC\,2852961 was also covered in three XMM-Newton slew observations (in 2006, 2013 and 2017), but in these short (45\,s) exposures the source was not detected. So we derived upper limits using the Rapid-XMM tool \citep{Ruiz2022}. The 3$\sigma$ upper limit count rates at the star's position correspond to X-ray luminosities of $1.65\times10^{32}$, $9.64\times10^{31}$, and $1.37\times10^{32}$\,erg\,s$^{-1}$ for the three slew observations, using the count-to-flux conversion factor from \citet{Freund2018}, which includes conversion to the ROSAT band for easier comparison with past ROSAT observations. The derived upper limits are consistent with the earlier RASS detection.

Data from the eROSITA All-Sky Survey are not publicly available, as KIC\,2852961 is located in the Russian part of the sky (eastern Galactic hemisphere). Other available X-ray all-sky surveys have too hard energy bands to be useful for typical stellar coronal emission.

Regarding the X-ray luminosities estimated above, we note that the adopted count-to-flux conversion factors were determined either using a small value for the interstellar hydrogen column density suitable for nearby stars \citep[$N_{\rm H}\sim10^{19}$\,cm$^{-2}$;][]{Huensch1996} or neglecting it \citep{Freund2018}. However, at the distance of KIC\,2852961 the interstellar hydrogen column density is larger as well, and thus the estimated X-ray luminosities are lower limits. From the interstellar extinction estimated in Sect.\,\ref{sect:stellarparams}, a hydrogen column density of $N_{\rm H}=(5.15\pm0.21)\times10^{20}$~cm$^{-2}$ can be derived \citep{2009MNRAS.400.2050G}. The unabsorbed (intrinsic) fluxes calculated from this $N_{\rm H}$ value using WebPIMMS\footnote{\url{https://heasarc.gsfc.nasa.gov/cgi-bin/Tools/w3pimms/w3pimms.pl}} are about a factor of 1.5 higher. The corrected ROSAT X-ray luminosity is therefore $(8.87\pm1.75)\times10^{31}$\,erg\,s$^{-1}$, accordingly $\log{(L_X/L_\mathrm{bol})=-3.47}$, while the aforementioned XMM-Newton 3$\sigma$ upper limits rise to $2.4\times10^{32}$, $1.4\times10^{32}$, and $2.0\times10^{32}$ erg\,s$^{-1}$,
respectively.

Due to the binary nature of KIC\,2852961, we also infer that the secondary star has a negligible contribution to the system's X-ray luminosity. As estimated above, the secondary is a mid-type M dwarf. At an age of $\sim$1.5\,Gyr, mid- to late-type M dwarfs are still in the saturated regime \citep{Engle2024}, which indicates that their X-ray emission can be up to around $10^{-3}L_\mathrm{bol}$. Using an estimated $L_\mathrm{bol}$ of 0.016\,$L_\odot$ for an $\sim$M4 dwarf, this gives a maximum value of $L_\mathrm{X}\sim6\times10^{28}$\,erg\,s$^{-1}$, three orders of magnitude lower than the observed X-ray emission, which is thus dominated by the giant as expected.

\subsection{What makes an active red giant produce superflares?}

X-ray luminosity is widely used as a proxy for the total magnetic energy and magnetic flux in the coronae of the Sun and late-type stars, including active red giants. If the occurrence of superflares were primarily controlled by the level of global magnetic activity, then we would expect red giant stars with higher coronal X-ray luminosity to be more prolific superflare producers. To test this proposal, we will use an RS\,CVn binary that is very similar to KIC\,2852961 in terms of its physical nature and magnetic activity characteristics.

XX\,Trianguli, a well-studied RS\,CVn type system \citep{2024NatCo..15.9986S,2025A&A...701A.103K}, shows remarkably similar global properties to KIC\,2852961, including orbital period, negligible orbital eccentricity with almost synchronous rotation, stellar radius, pronounced spottedness, high level of coronal activity ($\log(L_{\rm X}/L_{\rm bol})=-3.43$) and even a mid-to-late dM companion. At the same time, even at similar inclination angles, the rotational variability of XX\,Tri is actually significantly stronger: over the last two decades, the characteristic peak-to-peak amplitudes in the $V$-band were reported to be typically 0.1-0.2\,mag, equivalent to a change of up to 20\% (although historical values as high as $\Delta V$$\approx$0.5-0.7\,mag have also been recorded). This is considerably larger than the maximum peak-to-peak \emph{Kepler} modulation of about 7\% for KIC\,2852961. Nevertheless, no superflare events similar to those of KIC\,2852961 are known from the photometric records of XX\,Tri, despite its extended photometric history. Unfortunately, the available TESS observations of XX\,Tri provide a limited basis for comparison. The MAST archive\footnote{\url{https://mast.stsci.edu/portal/Mashup/Clients/Mast/Portal.html}} currently contains three TESS light curves for XX\,Tri from Sectors 18, 58, and 85, each covering approximately one rotation cycle with peak-to-peak amplitudes of 13\% and 9\%–9\%, respectively. However, the flare activity apparent in these data is indeed moderate: only few (at most four) low-amplitude events are visible, with no bursts comparable to the superflares observed on KIC\,2852961. It is a fact that the TESS data for XX\,Tri are far less extensive and temporally less continuous than the \emph{Kepler} and TESS observations of KIC\,2852961. Therefore, the absence of superflares in the XX\,Tri data does not provide a sufficiently solid basis for a quantitative comparison. However, the significantly stronger rotational modulation of XX\,Tri--combined with its presumably weak flare activity--suggests that the large-amplitude photometric variability caused by starspots may not, in itself, be sufficient to explain the extreme flare activity of KIC\,2852961.

Interestingly, the two giants differ in one key observable dynamo parameter, namely surface shear rate relative to rotation (i.e., $\alpha_{\rm DR}$=$\Delta\Omega/\Omega$), which is substantially stronger for KIC\,2852961 ($\alpha_{\rm DR}$$\sim$0.05) compared to XX\,Tri ($\alpha_{\rm DR}$=0.014$\pm$0.003; \citealt{2025A&A...701A.103K}). Although these shear values were determined using different methods, and thus carry different systematic uncertainties, the observed contrast is nevertheless consistent with the possibility that stronger surface shear contributes significantly to the generation and storage of magnetic energy. This may explain the different flare activities, because larger surface shear may indicate a more efficient transfer of rotational kinetic energy into non-potential magnetic fields, allowing greater amounts of free magnetic energy to accumulate before being released in superflares. In summary, it is possible that magnetic field complexity and shear, rather than the level of magnetic activity per se, are key components in controlling superflare activity in magnetically active RS\,CVn binaries \citep[in this regard, see also][]{2018ARep...62..513K}. This is also consistent with the conclusion drawn by \citet{2000A&A...361..614P}, that differential rotation is crucial for the X-ray emission level in the early post-main sequence evolutionary phase of low mass (\mbox{$\sim$1-3\,$M_{\odot}$)} stars.

\begin{figure}[thb]
    \centering
    \includegraphics[width=0.9\columnwidth]{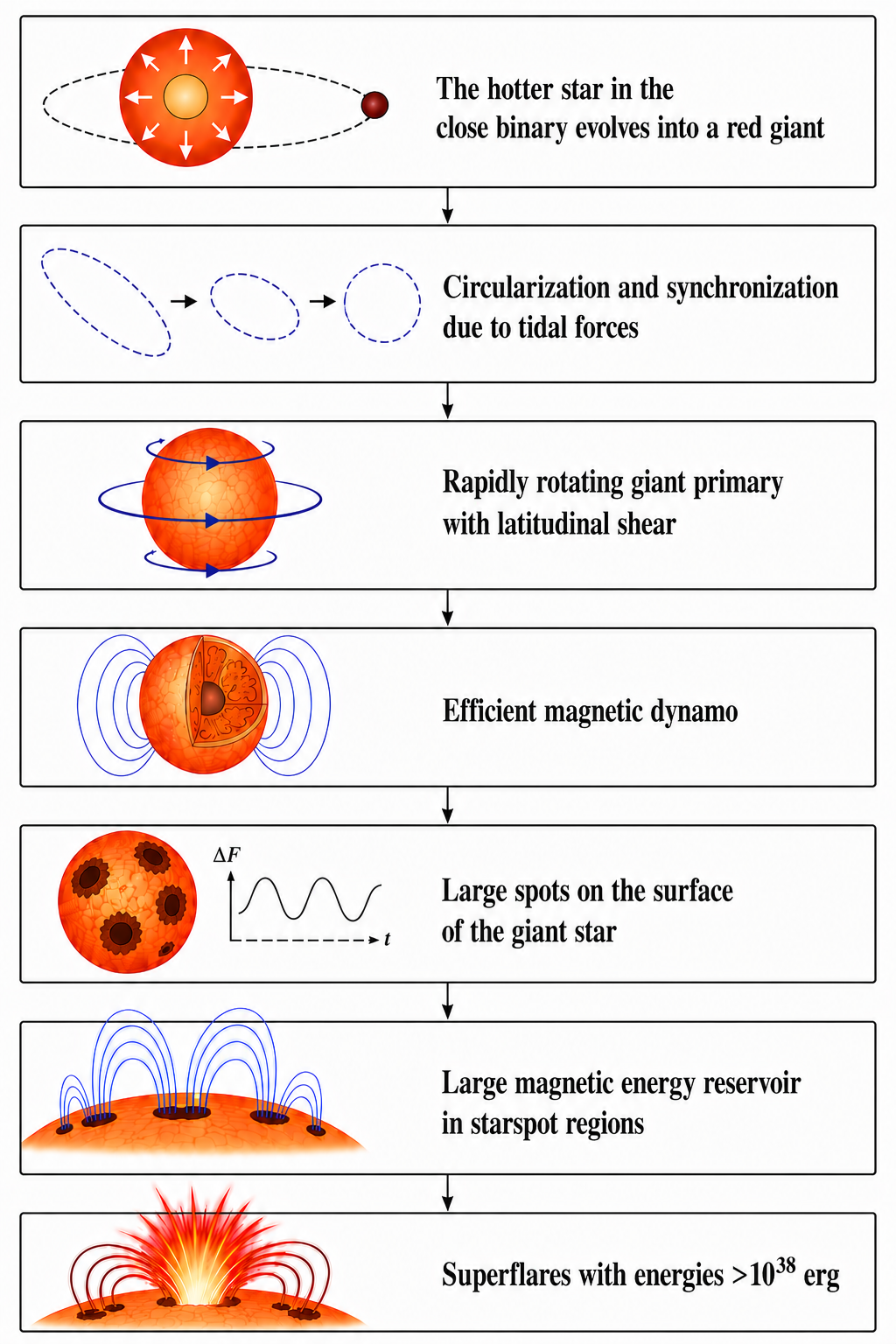}
    \caption{Hypothesized scenario leading to the formation of superflares in KIC\,2852961.}
    \label{fig:chart}
\end{figure}

The largest amplitudes of the rotational light modulation of KIC\,2852961 indicate a significantly non-axisymmetric spotted surface with large magnetically active regions capable of storing significant free magnetic energy. This, together with a moderate, but not negligible surface differential rotation provides favorable conditions for the build-up and subsequent release of magnetic energy, that is, the most energetic superflares observed in the \emph{Kepler} field. This supports a close connection between the extent and configuration of the spotted surface, surface differential rotation as a source of magnetic stress, magnetic energy storage, and the occurrence of energetic superflares. To summarize, Fig.\,\ref{fig:chart} shows a schematic illustration of the proposed evolutionary and magnetic-activity scenario for KIC\,2852961.

\section{Summary and conclusions}\label{Sect_sum}

In this study, we revisited KIC\,2852961, the red giant star producing the most powerful flares in the \emph{Kepler} field. We were curious about what physical characteristics lead to its outstanding flare activity. We have summarized our results leading to the final conclusion in the following points:

   \begin{enumerate}
      \item The new RV measurements, together with the improved Gaia DR-3 parallax, enabled a reliable orbital characterization of KIC\,2852961, leading to significantly improved constraints on both the stellar and system parameters.
      \item The near-circular orbit of KIC\,2852961 provides clear evidence that the expansion of the primary star and the deepening of its convective envelope have greatly enhanced tidal dissipation and accelerated orbital circularization, consistent with the estimated age of the system.
      \item Our time-series two-spot model was able to track the vivid brightness changes well, while also suggesting the presence of surface differential rotation. However, we found no correlation between spot phases and the occurrence of flares.
      \item TESS observations further strengthen the link between large spot-induced rotational modulation and more frequent high-energy superflares.
      \item The primary component of KIC\,2852961 exhibits a moderate (most likely) solar-like differential rotation with an estimated relative surface shear parameter of $\alpha_{\rm DR}$$\sim$0.05, demonstrating that tidal synchronization has been highly effective but has not completely suppressed rotational shear within the deep convective envelope.
      \item The value of $\log L_{\rm X}=31.95$\,[erg\,s$^{-1}$] derived from RASS together with $\log(L_{\rm X}/L_{\rm bol})=-3.47$ places KIC\,2852961 among the most magnetically active RS\,CVn giants.
      \item The flare occurrence rate exhibits no detectable dependence on orbital phase, indicating that the observed flares are governed primarily by the intrinsic magnetic activity of the active giant component rather than by phase-dependent interactions with the secondary star.
      \item The flare-frequency distribution, together with the expected increase in the available magnetic-energy reservoir with spot-induced rotational modulation, provides a physical context for the observed association of larger rotational modulation amplitudes of KIC\,2852961 with both higher characteristic flare energies and an increased (super)flaring rate. The latter connection is further supported by empirical evidence for an activity-dependent FFD normalization in other active stars.
    \end{enumerate}

With all these results, we conclude that the amplification and long-term efficiency of the magnetic dynamo behind the extraordinary superflare activity of KIC\,2852961 is a consequence of tidal interactions occuring at an earlier evolutionary stage of the close binary system, when the hotter star entered the red giant state. Although the mid-dM companion played a fundamental role in the evolution of the close binary and thus in maintaining the rapid rotation of the K giant, we found no evidence that the secondary star directly triggers the flare events, despite the small separation. Instead, the observed superflares are controlled by the intrinsic magnetic properties associated with the rapidly and differentially rotating deep turbulent-convective envelope of the active giant. We found that the differential rotation of the red giant is somewhat constrained by tidal forces, yet still efficient enough to support the dynamo and thus the formation of large starspots. Besides, differential rotation not only contributes to the generation of large-scale magnetic flux (i.e., large starspots), but also continuously shears and twists coronal magnetic structures, thereby helping to produce superflares.

\begin{acknowledgements}
We greatly appreciate the helpful advice provided by the anonymous Referee, which enabled our paper to be significantly improved. ZsK, BS, LK, KO, PO, ML, KV and RG would like to thank the Austrian-Hungarian Action Foundation for its financial support over many years, which has enabled continuous collaboration between researchers from both countries. We are particularly grateful for the grant 117öu4 running in 2024-2025, which contributed to the creation of this publication. Part of this work was supported by the Hungarian National Research, Development and Innovation Office grant KKP-143986. This research was funded in whole, or in part, by the Austrian Science Fund (FWF) 10.55776/I5711, 10.55776/P37256, 10.55776/PAT4657624. On behalf of the \textit{"Looking for stellar CMEs on different wavelengths"} project we are grateful for the possibility of using HUN-REN Cloud \cite{MTACloud} which helped us achieve the results published in this paper.

\end{acknowledgements}

\bibliographystyle{aa} %
\bibliography{bib.bib} %

\clearpage
\begin{appendix}

\section{Observing log of spectroscopic data}\label{appA}

\vspace{0.6cm}

\begin{center}
\tablehead{\hline\hline\noalign{\smallskip}%
HJD & ${\varv}_{\rm r}$ [km\,s$^{-1}$] & exp. time [s] & S/N \\
\noalign{\smallskip}\hline\noalign{\smallskip}}
\topcaption{Observing log for RV data from the 1-m RCC telescope at Piszkéstető, Hungary.}\label{Table:piszkes_RV}
\begin{supertabular}{crcr}
\tabletail{\noalign{\smallskip}\hline\noalign{\smallskip}}%
\tablelasttail{\noalign{\smallskip}\hline\noalign{\smallskip}}%
2458943.49806 & $23.543 \pm 0.366 $ &    3600 &      25 \\
2458943.54304 & $24.240 \pm 0.386 $ &    3600 &      20 \\
2458943.58590 & $24.512 \pm 0.376 $ &    3600 &      34 \\
2458944.53023 & $24.629 \pm 0.360 $ &    3600 &      45 \\
2458944.57293 & $24.314 \pm 0.345 $ &    3600 &      36 \\
2458944.48783 & $24.032 \pm 0.363 $ &    3600 &      36 \\
2458946.61765 & $24.200 \pm 0.873 $ &    1800 &      17 \\
2458947.52772 & $23.685 \pm 0.830 $ &    3600 &      14 \\
2458947.57032 & $24.106 \pm 0.574 $ &    3600 &      15 \\
2458948.61191 & $23.257 \pm 0.540 $ &    1800 &      25 \\
2458948.58588 & $22.689 \pm 0.690 $ &    1800 &      16 \\
2458948.55097 & $23.995 \pm 0.598 $ &    3600 &      29 \\
2458971.47814 & $12.003 \pm 0.412 $ &    1800 &      30 \\
2458971.39269 & $11.565 \pm 0.466 $ &    1800 &      27 \\
2458971.49912 & $11.840 \pm 0.386 $ &    1800 &      33 \\
2458971.41365 & $11.171 \pm 0.500 $ &    1800 &      18 \\
2458971.52005 & $12.321 \pm 0.389 $ &    1800 &      25 \\
2458971.43460 & $11.830 \pm 0.488 $ &    1800 &      30 \\
2458971.54105 & $12.068 \pm 0.374 $ &    1800 &      33 \\
2458971.45551 & $11.718 \pm 0.425 $ &    1800 &      24 \\
2458971.56198 & $12.616 \pm 0.418 $ &    1800 &      27 \\
2458972.57288 & $15.129 \pm 0.472 $ &    1800 &      15 \\
2458972.40342 & $14.578 \pm 0.453 $ &    3600 &      19 \\
2458972.44514 & $14.544 \pm 0.379 $ &    3600 &      43 \\
2458972.48687 & $14.676 \pm 0.358 $ &    3600 &      26 \\
2458972.52859 & $14.720 \pm 0.347 $ &    3600 &      32 \\
2458973.40070 & $15.777 \pm 0.512 $ &    3600 &      12 \\
2458973.44241 & $17.191 \pm 0.445 $ &    3600 &      18 \\
2458973.48413 & $17.018 \pm 0.401 $ &    3600 &      37 \\
2458973.52584 & $17.154 \pm 0.377 $ &    3600 &      23 \\
2458973.56756 & $18.380 \pm 0.413 $ &    3600 &      19 \\
2458974.46445 & $18.492 \pm 0.307 $ &    3600 &      59 \\
2458974.50931 & $18.493 \pm 0.301 $ &    3600 &      83 \\
2458974.42254 & $18.624 \pm 0.328 $ &    3600 &      54 \\
2460769.60725 & $3.736 \pm 0.428 $ &    3600 &      10 \\
2460774.48745 & $-0.613 \pm 0.300 $ &    3600 &      16 \\
2460774.52915 & $-0.712 \pm 0.300 $ &    3600 &      18 \\
2460777.54021 & $1.040 \pm 0.314 $ &    3600 &      50 \\
2460777.58193 & $1.345 \pm 0.316 $ &    3600 &      33 \\
2460778.52886 & $3.158 \pm 0.325 $ &    3600 &      82 \\
2460778.57556 & $3.088 \pm 0.320 $ &    3600 &      76 \\
2460899.40243 & $19.410 \pm 0.321 $ &    3600 &      25 \\
2460900.45233 & $15.471 \pm 0.300 $ &    3600 &      58 \\
2460901.44674 & $14.497 \pm 0.333 $ &    3600 &      20 \\
2460923.32816 & $19.501 \pm 0.312 $ &    1800 &      50 \\
2460923.34906 & $19.768 \pm 0.322 $ &    1800 &      55 \\
2460923.29628 & $19.367 \pm 0.312 $ &    3600 &      98 \\
2460923.44955 & $20.304 \pm 0.300 $ &    3600 &      83 \\
2460923.49281 & $20.320 \pm 0.311 $ &    3600 &      91 \\
2460923.53456 & $20.306 \pm 0.312 $ &    3600 &      68 \\
2460925.42962 & $23.462 \pm 0.329 $ &    3600 &      56 \\
2460925.47134 & $23.487 \pm 0.333 $ &    3600 &      57 \\
2460925.51589 & $23.658 \pm 0.342 $ &    3600 &      51 \\
2460927.42942 & $24.516 \pm 0.344 $ &    3600 &      34 \\
2460927.47244 & $24.547 \pm 0.324 $ &    3600 &      45 \\
2460927.51455 & $24.464 \pm 0.339 $ &    3600 &      49 \\
2460928.38243 & $24.421 \pm 0.344 $ &    3600 &      37 \\
2460951.31706 & $7.057 \pm 0.389 $ &    3600 &      17 \\
2460951.36778 & $6.958 \pm 0.338 $ &    3600 &      37 \\
2460951.41019 & $7.145 \pm 0.336 $ &    3600 &      30 \\
2460951.45241 & $7.680 \pm 0.343 $ &    3600 &      25 \\
2460952.26928 & $7.690 \pm 0.309 $ &    3600 &      60 \\
2460952.31111 & $7.916 \pm 0.308 $ &    3600 &      66 \\
2460952.35281 & $8.371 \pm 0.306 $ &    3600 &      67 \\
2460952.39469 & $8.514 \pm 0.314 $ &    3600 &      56 \\
2460952.43723 & $8.784 \pm 0.330 $ &    3600 &      41 \\
2460953.23219 & $10.887 \pm 0.300 $ &    3600 &      82 \\
2460956.32625 & $17.506 \pm 0.301 $ &    3600 &      72 \\
2460957.39291 & $19.604 \pm 0.345 $ &    3600 &      30 \\
2460979.23383 & $-1.228 \pm 0.300 $ &    3600 &     114 \\
2460979.27565 & $-1.118 \pm 0.300 $ &    3600 &      77 \\
2460979.32090 & $-0.894 \pm 0.300 $ &    3600 &      73 \\
2460979.36618 & $-1.183 \pm 0.325 $ &    3600 &      41 \\
2460979.40820 & $-1.000 \pm 0.366 $ &    3600 &      33 \\
2460981.24893 & $-0.019 \pm 0.304 $ &    3600 &      53 \\
2460981.29261 & $-0.564 \pm 0.305 $ &    3600 &      89 \\
2460981.33445 & $-0.554 \pm 0.307 $ &    3600 &      70 \\
2460981.38120 & $-0.666 \pm 0.346 $ &    3600 &      48 \\
2460984.22529 & $4.514 \pm 0.315 $ &    3600 &      85 \\
2460984.26701 & $4.284 \pm 0.317 $ &    3600 &      76 \\
2460984.30925 & $4.439 \pm 0.320 $ &    3600 &      70 \\
2460984.35362 & $4.819 \pm 0.325 $ &    3600 &      60 \\
2460984.39625 & $4.867 \pm 0.323 $ &    3600 &      93 \\
2460985.21821 & $5.285 \pm 0.306 $ &    3600 &      81 \\
2460985.25994 & $5.557 \pm 0.310 $ &    3600 &      71 \\
2460985.30204 & $5.828 \pm 0.304 $ &    3600 &      99 \\
2461155.53274 & $4.477 \pm 0.317 $ &    3600 &      55 \\
2461156.51003 & $7.218 \pm 0.335 $ &    3600 &      56 \\
2461160.55554 & $16.858 \pm 0.393 $ &    3600 &      14 \\
2461259.47149 & $ 7.265 \pm 0.312 $ &    3600 &      57 \\
2461260.52999 & $ 9.297 \pm 0.339 $ &    3600 &      21 \\
2461261.36816 & $10.941 \pm 0.322 $ &    3600 &      23 \\
2461262.37668 & $13.200 \pm 0.300 $ &    3600 &      71 \\
2461263.36841 & $14.561 \pm 0.300 $ &    3600 &      75 \\
2461264.35910 & $17.900 \pm 0.317 $ &    3600 &      36 \\
2461265.35505 & $18.947 \pm 0.333 $ &    3600 &      35 \\
2461266.35538 & $22.687 \pm 0.316 $ &    3600 &      78 \\
2461267.35177 & $23.093 \pm 0.300 $ &    3600 &     104 \\
2461268.36436 & $24.185 \pm 0.300 $ &    3600 &      83 \\
2461269.33387 & $23.840 \pm 0.303 $ &    3600 &      90 \\
2461272.36322 & $23.246 \pm 0.300 $ &    3600 &      64 \\

\end{supertabular}

\newpage

\vspace{0.6cm}
\tablehead{\hline\hline\noalign{\smallskip}%
HJD & ${\varv}_{\rm r}$ [km\,s$^{-1}$] & exp. time [s] & S/N \\
\noalign{\smallskip}\hline\noalign{\smallskip}}
\topcaption{Observing log for RV data from the STELLA robotic observatory at Tenerife, Spain.}\label{Table:STELLA_RV}
\begin{supertabular}{crcr}
\tabletail{\noalign{\smallskip}\hline\noalign{\smallskip}}%
\tablelasttail{\noalign{\smallskip}\hline\noalign{\smallskip}}%
2460626.37911   &  $17.367 \pm 0.012 $   &  1800 & 35.0\\
2460645.37646   &  $10.091 \pm 0.024  $  &  1800 & 33.6\\
2460715.74544   &  $14.730 \pm 0.017  $  &  1800 & 26.6\\
2460932.41528   &  $20.430 \pm 0.008  $  &  1800 & 30.2\\
2460940.56993   &  $3.430 \pm 0.006  $  &  1800 & 29.7\\
2460946.43969   &  $-0.489 \pm 0.006  $ &  1800 & 35.0\\
2460950.42815   &  $4.647 \pm 0.018  $ &  1800 & 30.6\\
2461083.73790   &  $-0.729 \pm 0.041 $  &  1800 & 18.5\\
2461214.40357   &  $2.760 \pm 0.015 $  &  1800 & 31.6\\
2461216.43810   &  $-0.085 \pm 0.011 $  &  1800 & 32.8\\
2461217.50948   &  $-0.882 \pm 0.017  $  &  1800 & 35.7\\
2461218.51177   &  $-1.212 \pm 0.022  $  &  1800 & 34.2\\
2461219.53950   &  $-1.053 \pm 0.014 $  &  1800 & 36.3\\
2461224.40910   &  $5.158 \pm 0.011  $  &  1800 &  28.9\\
2461225.50811   &  $7.544 \pm 0.006  $  &  1800 &  36.5\\
2461235.60726   &  $24.528 \pm 0.021 $  &  1800 &  29.0\\
2461236.60981   &  $24.201 \pm 0.010 $  &  1800 & 23.9\\
2461237.66688   &  $23.804 \pm 0.015 $  &  1800 & 20.9\\
2461247.54995	&  $4.661 \pm 0.008	 $  &  1800 & 27.0\\
2461248.56679   &  $2.622 \pm 0.013  $  &  1800 & 25.9\\
2461249.56907   &  $1.386 \pm 0.018  $  &  1800 & 24.5\\
2461259.39176   &   $6.596 \pm 0.008 $  &  1800 & 25.9\\
2461262.63224   &  $14.248 \pm 0.005 $  &  1800 & 15.4\\
2461263.64885   &  $16.775 \pm 0.007 $  &  1800 & 27.8\\
2461264.67489   &  $18.784 \pm 0.011 $  &  1800 & 27.1\\

\end{supertabular}

\end{center}

\newpage
\onecolumn
\section{Posterior distribution of the orbital solution}\label{app_posteriors}

\begin{figure*}[h]
    \centering
    \includegraphics[width=\columnwidth]{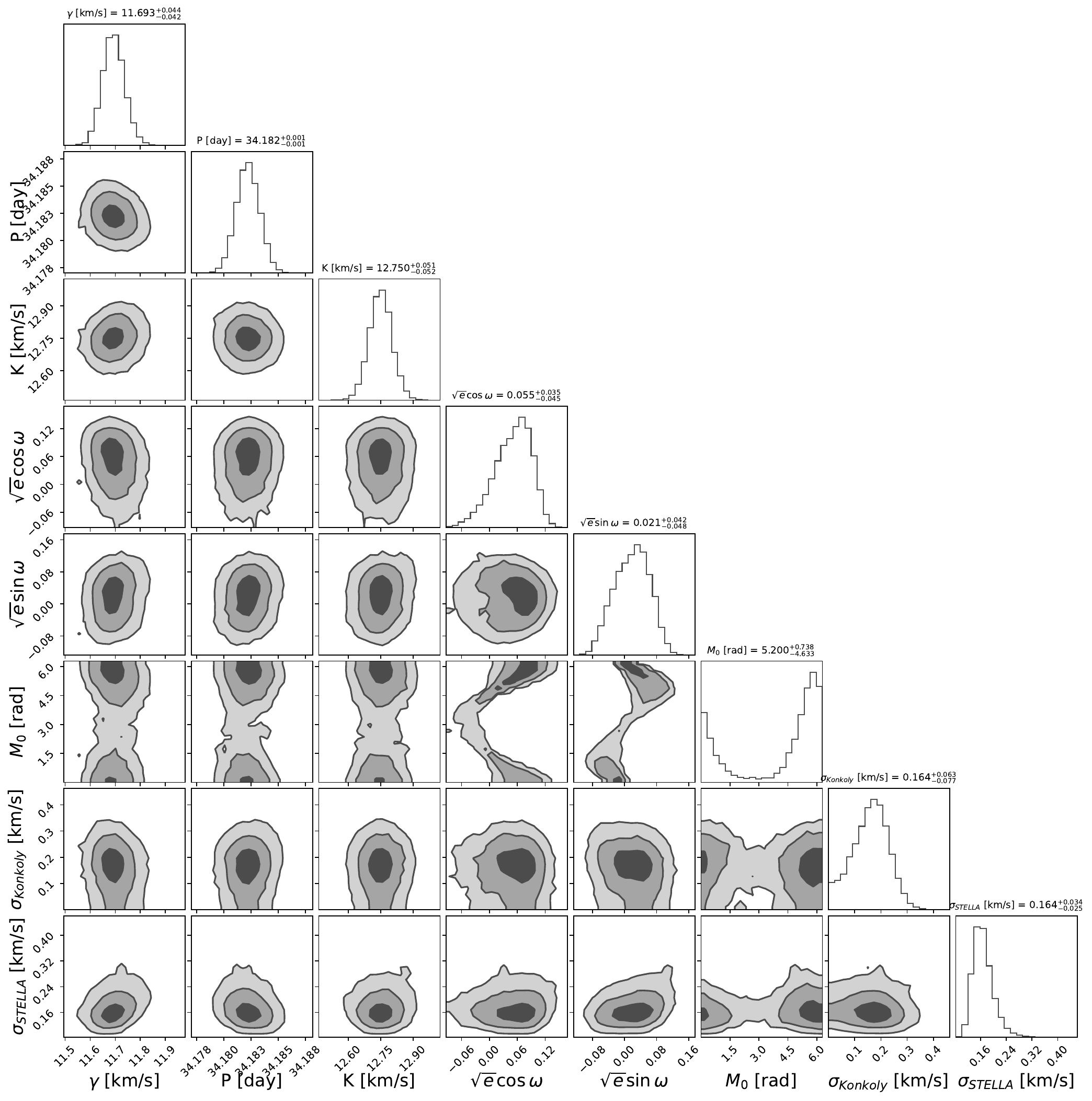}
    \caption{Corner plot of the posterior probability distribution of the orbital parameters from the MCMC fit.}
    \label{fig:RV_posteior}
\end{figure*}

\clearpage
\section{TESS light curves}
\label{sect:appendix_TESS_lc}

\begin{figure*}[hb]
    \centering    \includegraphics[width=0.9\columnwidth]{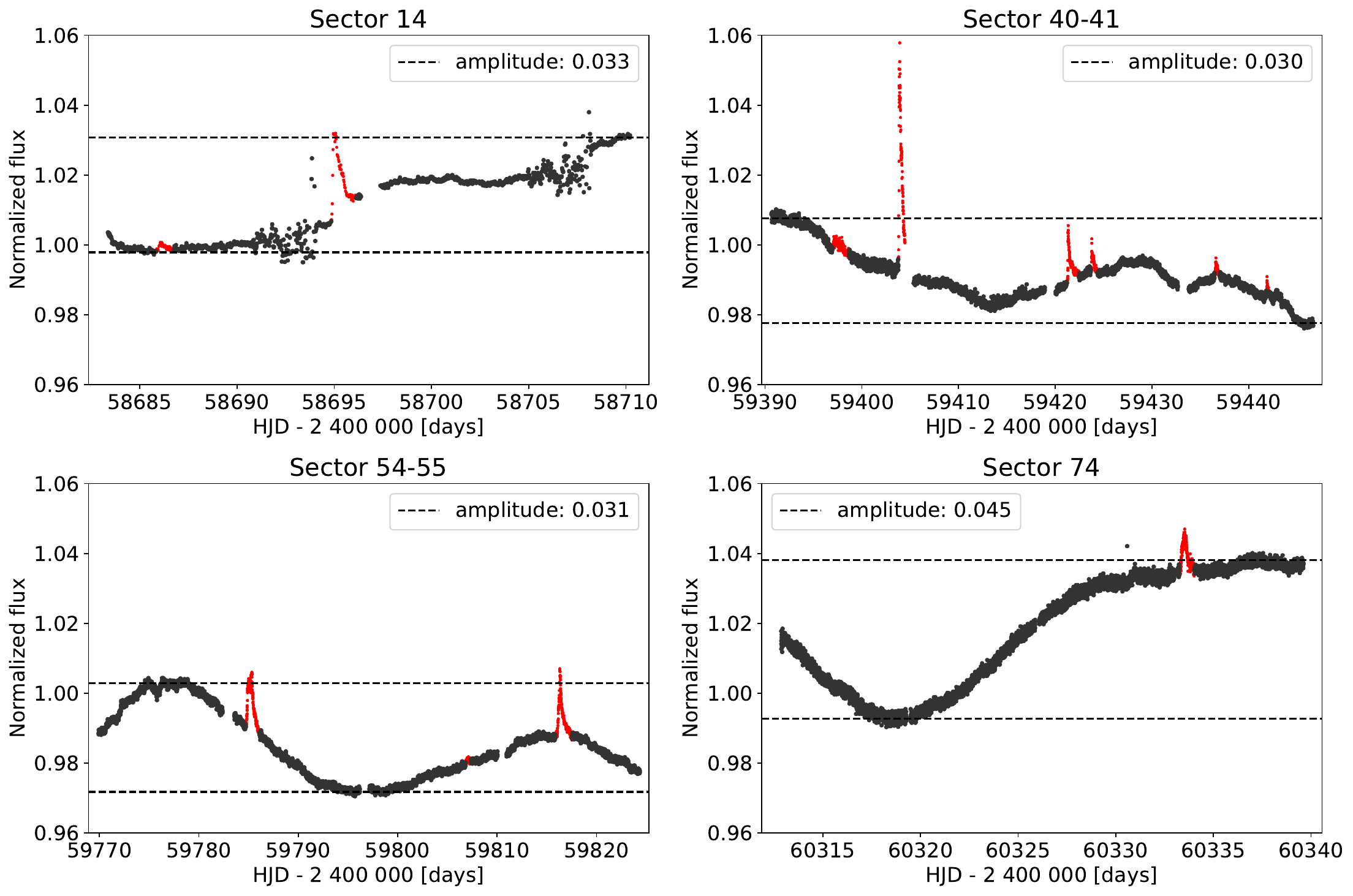}
    \caption{The available TESS data with the 12 identified flares marked in red. The dashed lines, which help with amplitude estimation, show the 2nd and 98th percentiles.}
    \label{fig:TESS_lc}
\end{figure*}

\clearpage
\section{Time-series spot modelling}
\label{sect:appendix_SML}

\begin{figure*}[h]
   \centering
   \includegraphics[width=0.85\columnwidth]{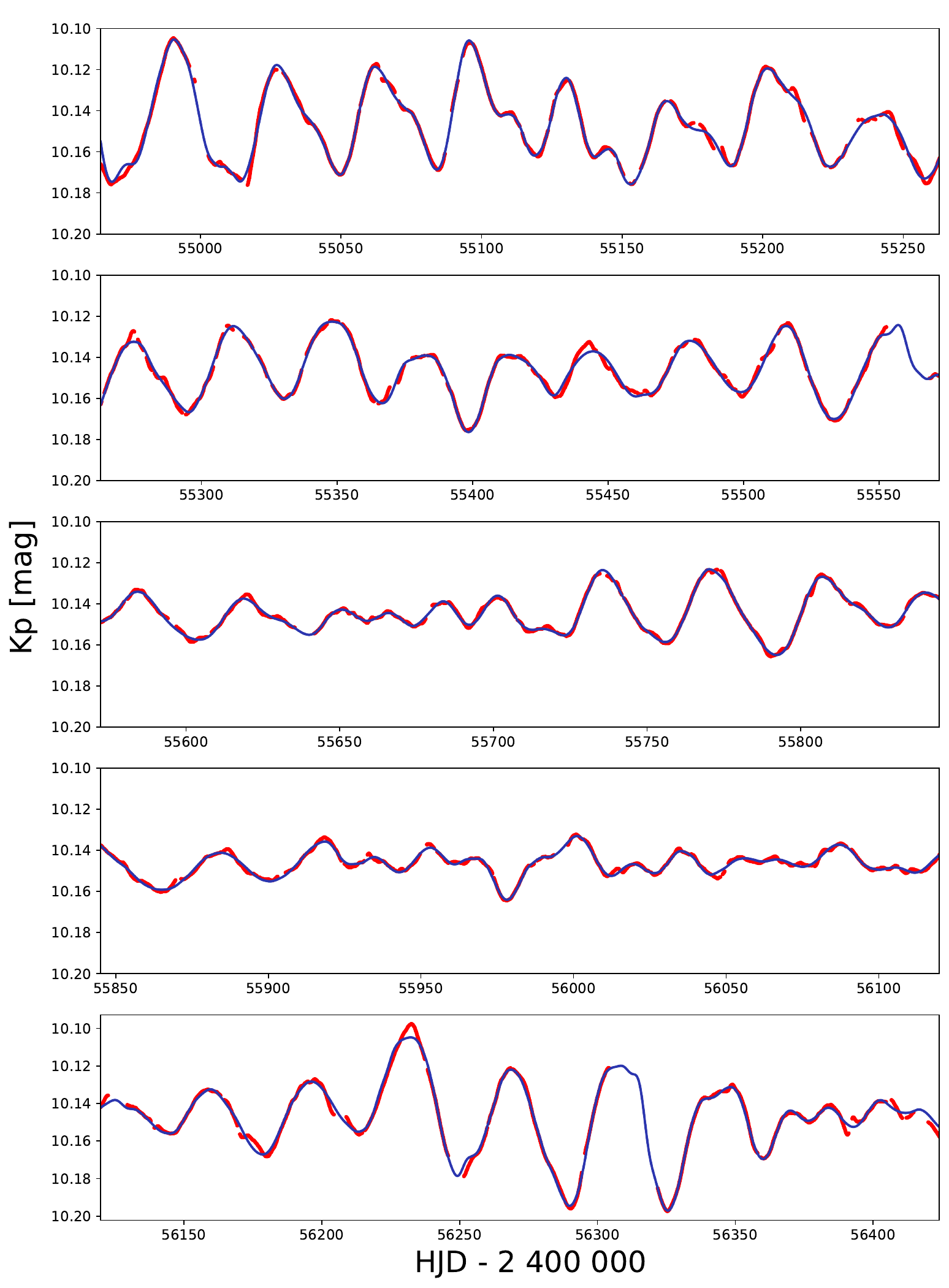}
   \caption{Fitting the full \emph{Kepler} light curve (red) with a time-series two-spot model (blue) with six free parameters (spot phases, latitudes and radii) using the SpotModel code.}
   \label{fig:attempt1_lc}
\end{figure*}

\begin{figure}[t]
   \centering
   \includegraphics[width=0.8\columnwidth]{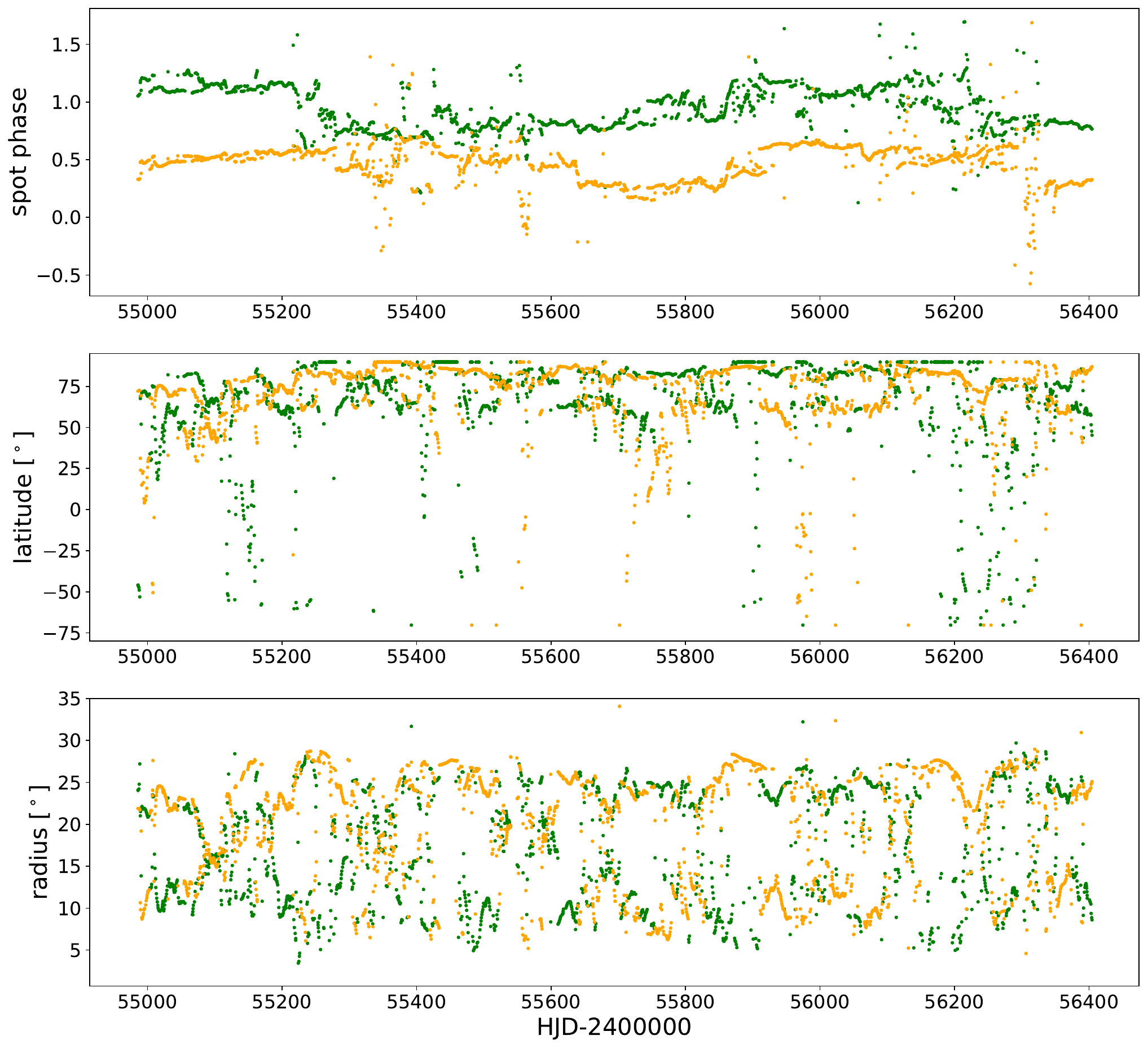}
   \caption{Spot coordinates from the time-series two-spot model (six-free-parameter fit) using the SpotModel code. The panels from top to bottom show the current spot phases, latitudes and radii, respectively. The coordinates of the two spots are represented in different colors (orange and green).}
   \label{fig:attempt1_spotpars}
\end{figure}

\begin{figure}[thb]
   \centering
   \includegraphics[width=0.42\columnwidth]{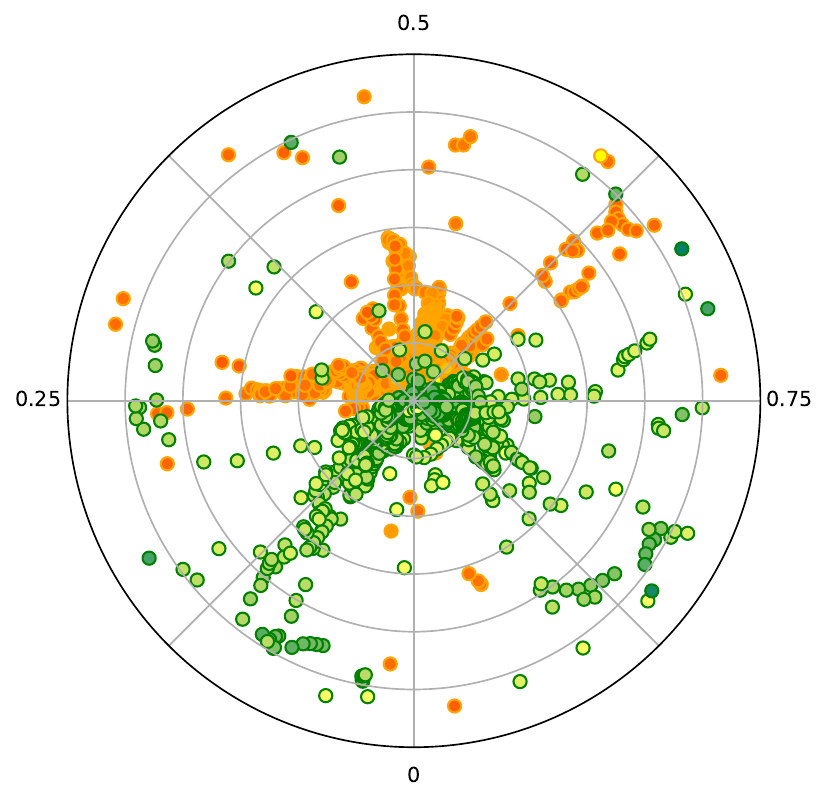}
   \caption{Pole-on diagram of
the spots' locations in the reference frame of the binary system using SpotModel with six free parameters. In this representation, concentric circles of increasing radius around the centre (visible pole) correspond to the latitude coordinates 60$^{\circ}$, 30$^{\circ}$, 0$^{\circ}$, $-$30$^{\circ}$, $-$60$^{\circ}$, $-$90$^{\circ}$. The positions of the spots, that is, phase-latitude ($\phi_i$; $\beta_i$) coordinate pairs, are indicated by circles. The color saturation of the circles indicate the sizes of the spots (the darker the color, the larger the spot).}
   \label{fig:attempt1_poleon}
\end{figure}

\begin{figure}[t]
   \centering
   \includegraphics[width=0.85\columnwidth]{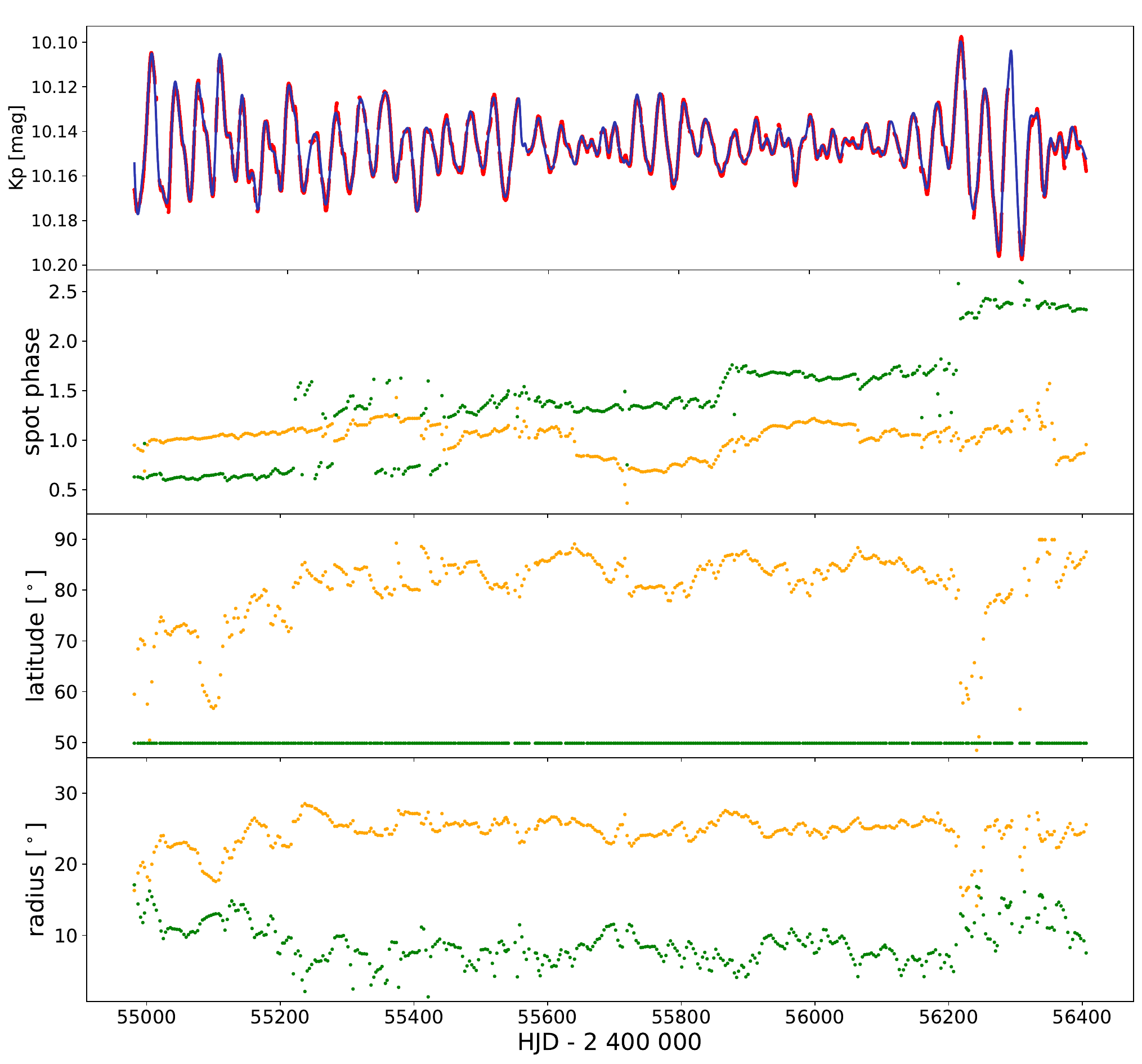}
   \caption{The result of our two-spot model using the SpotModel code, in which the latitude coordinate of one spot was fixed at $\beta$=50$^{\circ}$. The top panel shows the time-series model fit (blue) of the full \emph{Kepler} light curve (red), while the panels below show the current spot coordinates (spot phases, latitudes and radii). The coordinates of the two spots are represented in different colors (orange and green). That is, this figure was essentially created under the same conditions as Fig.\,\ref{fig:SML}, only the modeling code was different.}

\label{fig:attempt2_SpotModel}
\end{figure}

\end{appendix}

\end{document}